\documentclass[twocolumn,aps,showpacs,preprintnumbers,lettersize]{revtex4}
\usepackage{amsfonts}
\usepackage{amsmath}
\usepackage{amssymb}
\usepackage{graphicx}

\begin{document}

\title{Quantum Monte Carlo Study of Strongly Correlated Electrons: \\
Cellular Dynamical Mean-Field Theory}
\author{B. Kyung$^{1}$, G. Kotliar$^{2}$, and A. -M. S. Tremblay$^{1}$}
\affiliation{$^{1}$D\'{e}partement de physique and Regroupement qu\'{e}b\'{e}cois sur les
mat\'{e}riaux de pointe, Universit\'{e} de Sherbrooke, Sherbrooke, Qu\'{e}%
bec, J1K 2R1, Canada \\
$^{2}$Physics Department and Center for Materials Theory, Rutgers
University,Piscataway, New Jersey 08855, USA}
\date{\today }

\begin{abstract}
We study the Hubbard model using the Cellular Dynamical Mean-Field Theory
(CDMFT) with quantum Monte Carlo (QMC) simulations. We present the
algorithmic details of CDMFT with the Hirsch-Fye QMC method for the solution
of the self-consistently embedded quantum cluster problem. We use the one-
and two-dimensional half-filled Hubbard model to gauge the performance of
CDMFT+QMC particularly for small clusters by comparing with the exact
results and also with other quantum cluster methods. We calculate
single-particle Green's functions and self-energies on small clusters to
study their size dependence in one- and two-dimensions. It is shown that in
one dimension, CDMFT with two sites in the cluster is already able to
describe with high accuracy the evolution of the density as a function of
the chemical potential and the compressibility divergence at the Mott
transition, in good agreement with the exact Bethe ansatz result. With
increasing $U$ the result on small clusters rapidly approaches that of the
infinite size cluster. Large scattering rate and a positive slope in the
real part of the self-energy in one-dimension suggest that the system is a
non-Fermi liquid for all the parameters studied here. In two-dimensions, at
intermediate to strong coupling, even the smallest cluster ($N_{c}=2\times 2$%
) accounts for more than $95\%$ of the correlation effect of the
infinite-size cluster in the single particle spectrum, suggesting that some
of the important problems in strongly correlated electron systems may be
studied highly accurately with a reasonable computational effort. Finally,
as an application that is sensitive to details of correlations, we show that
CDMFT+QMC can describe spin-charge separated Luttinger liquid physics in
one dimension. The spinon and holon branches appear only for sufficiently
large system sizes.
\end{abstract}

\maketitle


\section{Introduction}

\label{section1}

Strongly correlated electron systems realized in organic conductors, heavy
fermion compounds, transition metal oxides, and more recently high
temperature superconductors continue to challenge our understanding. Various
anomalous behaviors observed in these materials cannot be well understood
within conventional theoretical tools based on a Fermi liquid picture or a
perturbative scheme. Because these intriguing features appear in a
non-perturbative regime, numerical methods have played a key role. Exact
diagonalization (ED) and Quantum Monte Carlo (QMC) simulations~\cite%
{Dagotto:1994} are amongst the most popular approaches. However, severe
limitations due to small lattice size in ED and a minus sign problem in QMC
at low temperatures make it difficult to extract reliable low-energy physics 
from these calculations.

Recently, alternative approaches~\cite%
{HTJPK:1998,SPP:2000,LK:2000,KSPB:2001,P:2003,MJPH:2004}, such as the
Dynamical Cluster Approximation, Cluster Perturbation Theory, the
Self-Energy Functional Approach and Cellular Dynamical Mean-Filed Theory
(CDMFT) have been developed and have already given some promising results.
Most of these quantum cluster methods generalize the single-site Dynamical
Mean Field Theory (DMFT)~\cite{GK:1992,Jarrell:1992,GKKR:1996} to
incorporate short-range spatial correlations explicitly. In fact the DMFT
has provided the first unified scenario for the long standing problem of the
Mott transition in the Hubbard model, completely characterizing the
criticality associated with this transition in infinite dimension or when
spatial correlations are negligible. In spite of its great success in
answering some of the challenging questions in strongly correlated electron
systems, its limitation has been also recognized in understanding low
dimensional electronic systems such as high temperature superconductors for
instance. In particular, the observed normal state pseudogap in underdoped
cuprates~\cite{TS:1999} is in sharp contrast with the prediction of DMFT in
which any slight doping into the half-filled band always leads to a Fermi
liquid. Many of the discrepancies are traced back to the neglect of
short-range correlations in DMFT. The main objective of these alternative
approaches is to describe short-range spatial correlations explicitly and to
study the physics that emerges. CDMFT has been recently applied to the
Hubbard model using ED as cluster solver at zero temperature, as we will
discuss later, and to the model for layered organic conductors with QMC as a cluster solver~\cite%
{PBK:2004}.

In this work we focus on CDMFT using the Hirsch-Fye~\cite{HF:1986} QMC
method to solve the cluster problem and to study its performance
particularly for small clusters in the one- and two-dimensional half-filled
Hubbard model. The method is benchmarked against exact results in one
dimension. Then we calculate single-particle Green's functions and
self-energies on small clusters to study their size dependence in one- and
two-dimensions. As an application of the approach that is particularly
sensitive to system size, we study the appearance of spin-charge separated
Luttinger liquid away from half-filling in one dimension.

This paper is organized as follows. In Sec.~\ref{section2} we review CDMFT.
In Sec.~\ref{section3} we present algorithmic details of the Hirsch-Fye QMC
method which is used to solve the self-consistently embedded quantum cluster
problem. In Sec.~\ref{section4} we present the CDMFT+QMC algorithm. In Sec.~%
\ref{section5} we benchmark the approach against exact results. Then in Sec. %
\ref{section_convergence} we show our results for size dependence of
one-particle quantities in the one-dimensional and two-dimensional
half-filled Hubbard models. The application to the Luttinger liquid appears
in Sec. \ref{section5-4}. Finally, in Sec.~\ref{section6}, we conclude our
present work, suggesting future applications of CDMFT.


\section{The Cellular Dynamical Mean Field Theory (CDMFT)}

\label{section2}

Throughout the paper, we will use the one- and two-dimensional half-filled
Hubbard model as an example, 
\begin{equation}
H=\sum_{\langle ij\rangle ,\sigma }t_{ij}c_{i\sigma }^{\dagger }c_{j\sigma
}+U\sum_{i}n_{i\uparrow }n_{i\downarrow }-\mu \sum_{i\sigma }c_{i\sigma
}^{\dagger }c_{i\sigma },  \label{eq10}
\end{equation}%
where $c_{i\sigma }^{\dagger }$ ($c_{i\sigma }$) are creation 
(annihilation) operators for electrons of spin $\sigma $, $n_{i\sigma
}=c_{i\sigma }^{\dagger }c_{i\sigma }$ is the density of $\sigma $ spin
electrons, $t_{ij}$ is the hopping amplitude equal to $-t$ for nearest
neighbors only, $U$ is the on-site repulsive interaction and $\mu $ is the
chemical potential controlling the electron density.

\begin{figure}[tbp]
\includegraphics[width=8.0cm]{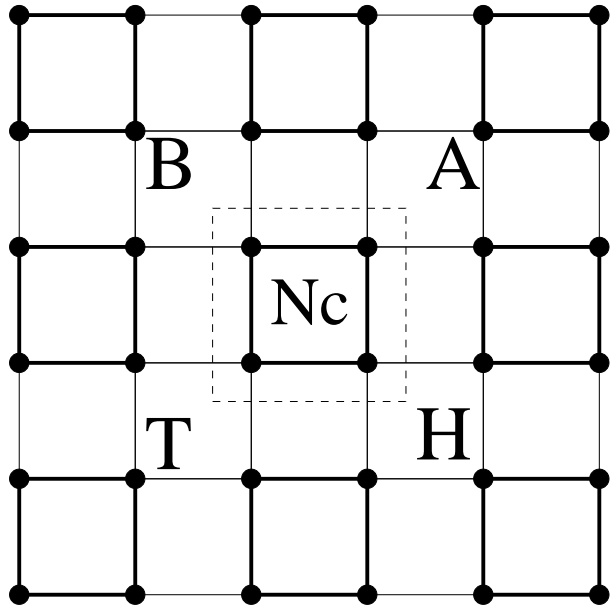}
\caption{CDMFT construction. The entire infinite lattice is tiled with
identical clusters of size $N_{c}$ in real space.}
\label{CDMFT.fig}
\end{figure}

The CDMFT~\cite{KSPB:2001} is a natural generalization of the single site
DMFT that treats short-range spatial correlations explicitly. Also, some
kinds of long-range order involving several lattice sites, such as $d$-wave
superconductivity, can be described in CDMFT and not in DMFT~\cite%
{KCCKSKT:2005}. In the CDMFT construction~\cite{KSPB:2001,BKK:2003} shown in
Fig.~\ref{CDMFT.fig}, the entire infinite lattice 
is tiled with identical clusters of size $N_{c}$. Degrees of freedom within
a cluster are treated exactly, while those outside the cluster are replaced
by a \textit{bath} of noninteracting electrons that is determined
self-consistently. This method~\cite{BKK:2003,CCKCK:2004} has already passed
several tests against some exact results obtained by Bethe ansatz and
density matrix renormalization group (DMRG) technique in one dimension. This
is where the DMFT or CDMFT schemes are expected to be in the worst case
scenario since DMFT itself is exact only in infinite dimension and
mean-field methods usually degrade as dimension is lowered. Nevertheless,
the CDMFT in conjunction with ED correctly predicts the divergence of the
compressibility at the Mott transition in one-dimension, a divergence that
is missed in the single-site DMFT.

We now recall the general procedure to obtain the self-consistency loop in
CDMFT in a manner independent of which method is used to solve the quantum
cluster problem. We also refer to Ref.~\cite{P:2003} for an alternate
derivation. The first CDMFT equation begins by integrating out the bath
degrees of freedom to obtain an $N_{c}\times N_{c}$ dynamical Weiss field $%
G_{0,\sigma }(i\omega _{n})$ (in matrix notation) where $i\omega _{n}$ is
the fermionic Matsubara frequency. This dynamical Weiss field is like the
Weiss field in a mean-field analysis of the Ising model. Because it contains
a full frequency dependence, it is dynamical instead of static and takes
care of quantum fluctuations beyond the cluster. The second CDMFT equation
defines the cluster self-energy from the cluster Green's function by solving
the quantum impurity problem and extracting $\Sigma (i\omega _{n})$ from 
\begin{equation}
\Sigma (i\omega _{n})=G_{0}^{-1}(i\omega _{n})-G_{c}^{-1}(i\omega _{n})\;.
\label{eq20}
\end{equation}%
To close the self-consistency loop, we obtain a new Weiss field using the
self-consistency condition 
\begin{eqnarray}
G_{0}^{-1}(i\omega _{n}) &=&\left( \frac{N_{c}}{(2\pi )^{d}}\int d\tilde{{\bf k}}%
\frac{1}{i\omega _{n}+\mu -t(\tilde{{\bf k}})-\Sigma (i\omega _{n})}\right) ^{-1}%
\kern-1em  \notag \\
&+&\Sigma (i\omega _{n})\;,  \label{eq30}
\end{eqnarray}%
where $d$ is a spatial dimension. Here $t(\tilde{{\bf k}})$ is the hopping matrix
for the superlattice with the wavevector $\tilde{{\bf k}}$ because of the
inter-cluster hopping. We go through the self-consistency loop until the old
and new Weiss fields converge within desired accuracy. Finally, after
convergence is reached, the lattice Green's function $G({\bf k},i\omega _{n})
$ is obtained using 
\begin{eqnarray}
G({\bf k},i\omega _{n}) &=&\frac{1}{N_{c}}\sum_{\mu \nu }e^{i{\bf k}\cdot (%
\vec{r}_{\mu }-\vec{r}_{\nu })}  \notag \\
&\times &\left[ \frac{1}{i\omega _{n}+\mu -t(\tilde{{\bf k}})-\Sigma (i\omega _{n})}%
\right] _{\mu \nu }\;,  \label{eq40}
\end{eqnarray}%
where $\Sigma (i\omega _{n})$ is the converged cluster self-energy, 
${\bf k}$ is any vector in the original Brillouin zone and $\mu \nu $ label cluster sites.
This last step differs~\cite{KKSTCK:2005} and improves
that proposed in Ref.~\cite{KSPB:2001}. See also Ref.~\cite{SK:2005}. The
lattice quantities such as the spectral function and the self-energy shown
in this paper are computed from this lattice Green's function.


\section{Quantum Monte Carlo Simulations}

\label{section3}

\subsection{Quantum Monte Carlo Method}

\label{section3-1}

In this section we present the algorithmic details of the Hirsch-Fye QMC
method~\cite{HF:1986,DS:2003} for the solution of the self-consistently
embedded quantum cluster problem. The basic principle of the QMC method can
be understood as a discretization of the quantum impurity model effective
action 
\begin{eqnarray}
S_{eff} &\rightarrow &\sum_{\mu \mu ^{\prime }\tau \tau ^{\prime }\sigma
}c_{\sigma }^{\dagger }(\mu \tau )G_{0,\sigma }^{-1}(\mu \mu ^{\prime },\tau
\tau ^{\prime })c_{\sigma }(\mu ^{\prime }\tau ^{\prime })  \notag \\
&+&U\sum_{\mu \tau }n_{\uparrow }(\mu \tau )n_{\downarrow }(\mu \tau ) \;,
\label{eq50}
\end{eqnarray}%
where the imaginary-time is discretized in $L$ slices $l=1,2,\cdots ,L$ of $%
\Delta \tau $, and the timestep $\Delta \tau $ is defined by $\beta =L\Delta
\tau $. Here $\beta =1/T$ is the inverse temperature in units where
Boltzmann's constant is unity. Throughout the paper $\Delta \tau =\sqrt{1/8tU%
}$ is used, unless otherwise specifically mentioned. This leads to a
systematic discretization error of order $\left( \Delta \tau \right) ^{2}$
which is a few percent. The remaining quartic term can be decoupled using a
discrete Hirsch-Hubbard-Stratonovich transformation~\cite{Hirsch:1983} 
\begin{equation*}
e^{-\Delta \tau Un_{\uparrow }n_{\downarrow }}=\frac{1}{2}e^{-\Delta \tau
U/2(n_{\uparrow }+n_{\downarrow })}\sum_{s=\pm 1}e^{\lambda s(n_{\uparrow
}-n_{\downarrow })} \;,
\end{equation*}%
where $\lambda =\cosh ^{-1}(e^{\Delta \tau U/2})$ and the discrete field $s$
is an Ising-like variable taking the values $\pm 1$. Performing this
transformation at every discrete space and imaginary-time point, we are led
to a quadratic action, and the partition function becomes, in a functional
integral representation, 
\begin{eqnarray}
Z &\propto &\sum_{s_{\mu l}=\pm 1}\int D[c^{\dagger },c]\exp \biggl \{%
-\sum_{\mu \mu ^{\prime }ll^{\prime }\sigma }c_{\sigma }^{\dagger }(\mu
l)G_{0,\sigma }^{-1}(\mu \mu ^{\prime },ll^{\prime })  \notag \\
&\times &c_{\sigma }(\mu ^{\prime }l^{\prime })  \notag \\
&+&\lambda \sum_{\mu l}s_{\mu l}(n_{\uparrow }(\mu l)-n_{\downarrow }(\mu l))%
\biggr \}  \notag \\
&=&\sum_{s_{\mu l}=\pm 1}\int D[c^{\dagger },c]\exp \biggl \{-\sum_{\mu \mu
^{\prime }ll^{\prime }\sigma }c_{\sigma }^{\dagger }(\mu l)G_{\sigma
,\{s\}}^{-1}(\mu \mu ^{\prime },ll^{\prime })  \notag \\
&\times &c_{\sigma }(\mu ^{\prime }l^{\prime })\biggr \}\;  \notag \\
&\propto &\sum_{s_{\mu l}=\pm 1}\prod_{\sigma }det(G_{\sigma ,\{s\}}^{-1})\;.
\label{eq70}
\end{eqnarray}%
The inverse propagator $G_{\sigma ,\{s\}}^{-1}(\mu \mu ^{\prime },ll^{\prime
})$ for a particular realization of the Ising spins $\{s\}$ is defined as 
\begin{equation}
G_{\sigma ,\{s\}}^{-1}(\mu \mu ^{\prime },ll^{\prime })=G_{0,\sigma
}^{-1}(\mu \mu ^{\prime },ll^{\prime })-\sigma \lambda s_{\mu l}\delta _{\mu
,\mu ^{\prime }}\delta _{l,l^{\prime }+1} \;,  \label{eq80}
\end{equation}%
where the antiperiodic delta function $\delta _{l,l^{\prime }+1}$~\cite%
{Comment:Deltafunction} is defined as $1$ if $l=l^{\prime }+1$ and $-1$ if $%
l=1$ and $l^{\prime }=L$.

The influence of the discrete field $s$ at each space-time point appears in $%
e^{V},$ a diagonal matrix with elements $e^{V_{\sigma ,\{s\}}}(\mu \mu
^{\prime },ll^{\prime })=e^{\sigma \lambda s_{\mu l}}\delta _{\mu ,\mu
^{\prime }}\delta _{l,l^{\prime }}.$ 
The Green functions $G$ and $G^{\prime }\,$that are characterized 
by $%
e^{V}$ and $e^{V^{\prime }}$ respectively are related by 
\begin{equation}
G^{\prime }=G+(G-1)(e^{V^{\prime }-V}-1)G^{\prime } \;. \label{eq90}
\end{equation}%
In fact Eq.~\ref{eq80} is a special case of Eq.~\ref{eq90} when all Ising
spins $\{s\}$ are turned off, which reduces $e^{V}$ to the unity matrix, and
when $e^{V^{\prime }}$ is expanded to linear order in $V^{\prime }$.


\subsection{Monte Carlo Simulation}

\label{section3-2}

In a Monte Carlo simulation, a local change in Ising spin configuration $%
s_{\mu l}\rightarrow s_{\mu l}^{\prime }$ is proposed and accepted with a
transition probability $W=p(s\rightarrow s^{\prime })/p(s^{\prime
}\rightarrow s)$. Since Ising spin configurations are generated with a
probability proportional to $\prod_{\sigma }det(G_{\sigma ,\{s\}}^{-1})$
according to Eq.~\ref{eq70}, the detailed balance property requires 
\begin{equation*}
\frac{p(s\rightarrow s^{\prime })}{p(s^{\prime }\rightarrow s)}=\frac{%
\prod_{\sigma }\det (G_{\sigma ,\{s^{\prime }\}}^{-1})}{\prod_{\sigma }\det
(G_{\sigma ,\{s\}}^{-1})}\;.
\end{equation*}%
Note that, as usual, when the determinant is negative, the absolute value of
the determinant is used as a weight and the sign becomes part of the
observable. In the case of a single spin flip, say $s_{\nu m}^{\prime
}=-s_{\nu m}$, the transition probability can be greatly simplified by
rearranging Eq.~\ref{eq90} as follows 
\begin{eqnarray}
G^{\prime } &=&A^{-1}G\;,  \notag \\
A &=&1+(1-G)(e^{V^{\prime }-V}-1)  \label{eq110}
\end{eqnarray}%
and by noting that%
\begin{eqnarray}
\det A_{\sigma } &=&A_{\sigma }(\nu \nu ,mm)\;,  \notag \\
&=&1+(1-G_{\sigma }(\nu \nu ,mm))  \notag \\
&\times &[e^{V_{\sigma }^{\prime }(\nu \nu ,mm)-V_{\sigma }(\nu \nu
,mm)}-1]\;.  \label{eq120}
\end{eqnarray}%
As a result, the transition probability $W=\prod_{\sigma }\det A_{\sigma }$
is given as a simple product of numbers with a computational effort of O(1).
Two popular algorithms have been used to compute an acceptance probability $%
AP$ 
\begin{eqnarray}
AP &=&\frac{W}{1+W} \;, \\
AP &=&\left\{ 
\begin{array}{ll}
1 & \mbox{if $W > 1$} \\ 
W & \mbox{otherwise}%
\end{array}%
\right. \;.  \label{eq130}
\end{eqnarray}%
They are, respectively, the heat bath and the Metropolis algorithms. If the
move $s_{\nu m}\rightarrow s_{\nu m}^{\prime }=-s_{\nu m}$ is accepted, 
then the propagator must be
updated by using Eq.~\ref{eq110} and Eq.~\ref{eq120} with a computational
burden of $N_{c}^{2}L^{2}$ 
\begin{eqnarray}
G^{\prime }(\mu \mu ^{\prime },ll^{\prime }) &=&G(\mu \mu ^{\prime
},ll^{\prime })+[G(\mu \nu ,lm)-\delta _{\mu ,\nu }\delta _{l,m}]  \notag \\
&\times &[e^{V^{\prime }(\nu \nu ,mm)-V(\nu \nu ,mm)}-1]  \notag \\
&\times &[A(\nu \nu ,mm)]^{-1}G(\nu \mu ^{\prime },ml^{\prime })\;.
\label{eq140}
\end{eqnarray}

We regularly recompute the propagator $G(\mu \mu ^{\prime },ll^{\prime })$
with Eq.~\ref{eq90} or Eq.~\ref{eq110} to compensate a possible
deterioration (due to round-off error) of $G(\mu \mu ^{\prime },ll^{\prime
}) $ which is generated by a sequence of updates with Eq.~\ref{eq140}. After
several hundreds of warmup sweeps through the discrete space and
imaginary-time points of the cluster, we make measurements for the Green's
function, density and other interesting physical quantities. We reduce the
statistical error by using all available symmetries. That includes the
point-group symmetries of the cluster, the translational invariance in
imaginary-time, the spin symmetry in the absence of magnetic long-range
order and the particle-hole symmetry at half-filling. Results of the
measurements are accumulated in bins and error estimates are made from the
fluctuations of the binned measurements provided that the bins contain large
enough measurements so that the bin-averages are uncorrelated. Finally the
maximum entropy method (MEM)~\cite{JG:1996,Allen:2005} is used to perform
the numerical analytical continuation of the imaginary-time Green's function.

Because QMC simulations are performed in imaginary-time and the CDMFT
equations (Eq.~\ref{eq20}, Eq.~\ref{eq30}, Eq.~\ref{eq40}) are given in
Matsubara frequencies, special care must be taken in making Fourier
transforms. The direct Fourier transform at a finite number of discrete
imaginary-time steps renders the Green's function $G_{c}(i\omega _{n})$ (Eq.~%
\ref{eq20}) a periodic function of $i\omega _{n}$ instead of having the
correct asymptotic behavior $G_{c}(i\omega _{n})\sim 1/i\omega _{n}$ at
large Matsubara frequencies. We used a spline interpolation scheme 
\begin{eqnarray}
G_{c}^{interpol}(\tau ) &=&\alpha _{i}+\beta _{i}(\tau -\tau _{i})+\gamma
_{i}(\tau -\tau _{i})^{2}  \notag \\
&+&\delta _{i}(\tau -\tau _{i})^{3}\mbox{  for }\tau _{i}<\tau <\tau _{i+1}
\label{eq150} \;,
\end{eqnarray}%
where the coefficients $\alpha _{i},\beta _{i},\gamma _{i},\delta _{i}$ are
analytically calculated from the original Green's function obtained in
imaginary-time. Then the piecewise integral is performed $\int d\tau
G_{c}^{interpol}(\tau )e^{i\omega _{n}\tau }$ to compute $G_{c}(i\omega
_{n}) $. In practice, we subtract a reference function $G^{\prime }(\tau )$
and add the corresponding $G^{\prime }(i\omega _{n})$ which is known exactly
and chosen to have the same asymptotic behavior as $G_{c}(i\omega _{n})$ 
\begin{eqnarray}
G_{c}(i\omega _{n}) &=&G^{\prime }(i\omega _{n})  \notag \\
&+&\int d\tau \lbrack G_{c}(\tau )-G^{\prime }(\tau )]e^{i\omega _{n}\tau
}\;.  \label{eq160}
\end{eqnarray}%
Thus errors in the spline interpolation scheme applied to the difference of
the two functions can be reduced significantly. Recently another scheme~\cite%
{OK:2002} was proposed to calculate the correct high frequency behavior by
exploiting additional analytic information about the moments of $G_{c}(\tau
) $. For more algorithmic details of QMC simulations see Refs.~\cite%
{GKKR:1996,JMHM:2001,KSHOPM:2005}.


\section{The CDMFT+QMC Algorithm}

\label{section4}

In this section we outline the CDMFT algorithm in conjunction with the
Hirsch-Fye QMC method. 
\begin{figure}[tbp]
\includegraphics[width=10.0cm]{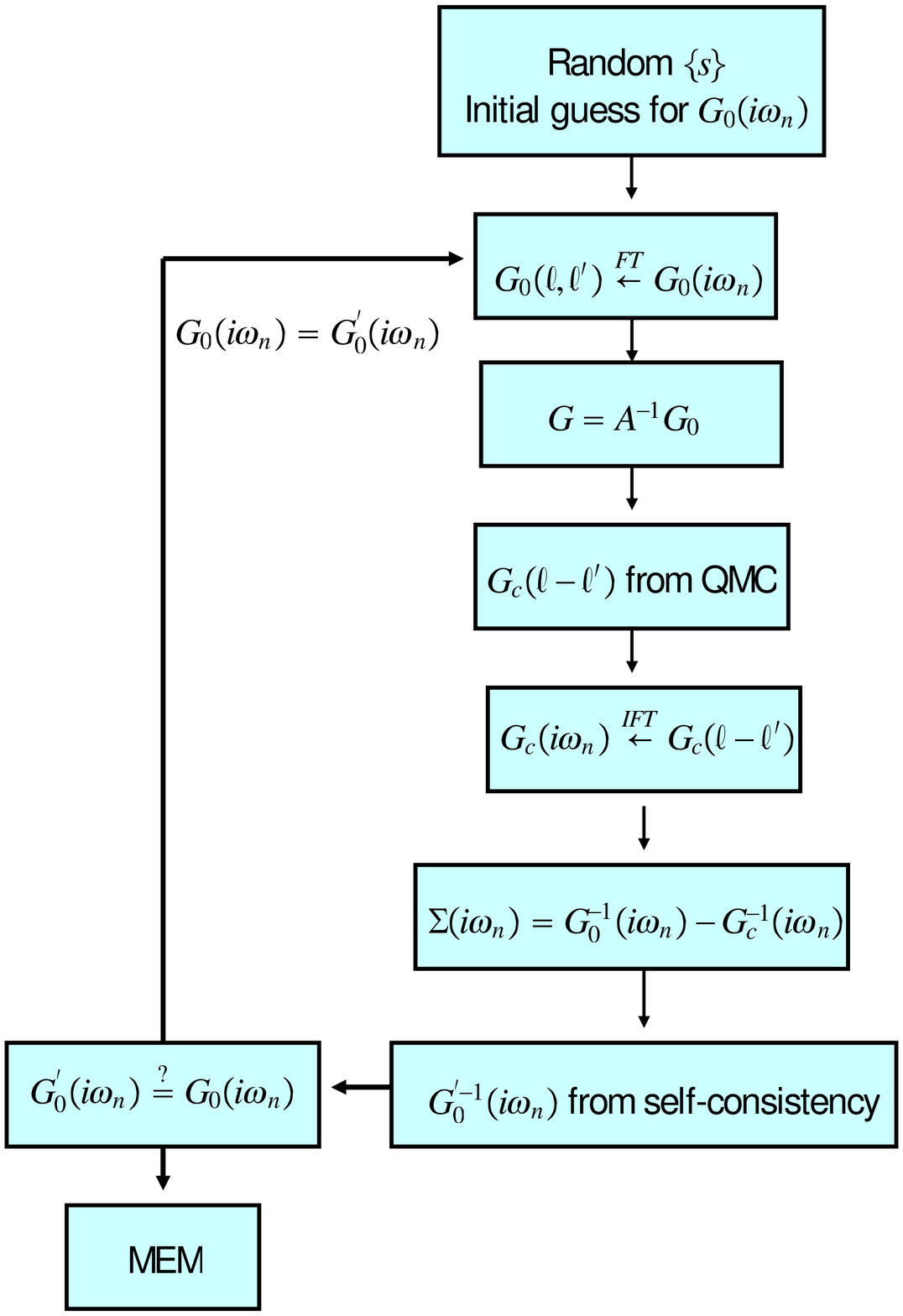}
\caption{Sketch of the CDMFT algorithm using QMC method. }
\label{CDMFT_chart.fig}
\end{figure}

(1) We start by generating a random Ising spin configuration and an initial
guess for the dynamical Weiss field $G_{0,\sigma }(\mu \mu ^{\prime
},i\omega _{n})$. The latter is usually taken as the non-interacting value. 
\newline
(2) The Weiss field is Fourier transformed (FT) to obtain $G_{0,\sigma }(\mu
\mu ^{\prime },ll^{\prime })$. \newline
(3) The propagator $G_{\sigma ,\{s\}}(\mu \mu ^{\prime },ll^{\prime })$ for
the Ising spin configuration with $s_{\mu l}=\pm 1$ is calculated by
explicit inversion of the matrix $A$ in Eq.~\ref{eq110} with $G$ replaced by 
$G_{0}$ in the latter equation. \newline
(4) From then on, configurations are visited using single spin flips. When
the change is accepted, the propagator is updated using Eq.~\ref{eq140}. 
\newline
(5) The physical cluster Green's function $G_{c,\sigma }(\mu \mu ^{\prime
},l-l^{\prime })$ is determined as averages of the configuration-dependent
propagator $G_{\sigma ,\{s\}}(\mu \mu ^{\prime },ll^{\prime }).$ The biased
sampling guarantees that the Ising spin configurations are weighted
according to Eq.~\ref{eq120}. \newline
(6) $G_{c,\sigma }(\mu \mu ^{\prime },l-l^{\prime })$ is inverse Fourier
transformed (IFT) by using a spline interpolation scheme (described in the
previous section) to obtain $G_{c,\sigma }(\mu \mu ^{\prime },i\omega _{n})$%
. \newline
(7) The cluster self-energy $\Sigma (\mu \mu ^{\prime },i\omega _{n})$ is
computed from the cluster Green's function using Eq.~\ref{eq20}. \newline
(8) A new dynamical Weiss field $G_{0,\sigma }^{\prime}(\mu \mu ^{\prime },
i\omega_{n})$ 
is calculated using the self-consistency condition Eq.~\ref{eq30}. 
\newline
(9) We go through the self-consistency loop (2)-(8) until the old and new
Weiss fields converge within desired accuracy. Usually in less than 10
iterations the accuracy reaches a plateau (for example, relative mean-square
deviation of $10^{-4}$ for $U=8,$ $\beta =5,$ or smaller for smaller
interaction strength). \newline
(10) After convergence is reached, the numerical analytical continuation is
performed with MEM on the data from the binned measurements. \newline
Figure~\ref{CDMFT_chart.fig} is a sketch of the CDMFT algorithm using the
QMC method.


\begin{figure}[tbp]
\includegraphics[width=8.0cm]{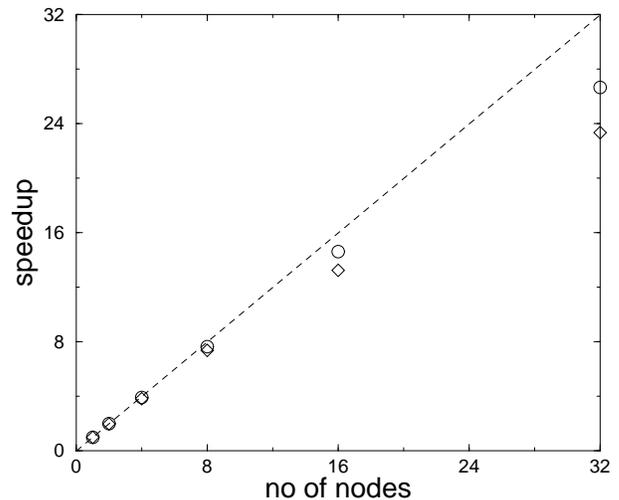}
\caption{Speedup versus the number of nodes using MPI. Circles and diamonds
represent speedup for the combined total number of measurements respectively
equal to 64,000 and 32,000.}
\label{MPI_SPEEDUP.fig}
\end{figure}
Fig.~\ref{MPI_SPEEDUP.fig} shows the speedup achieved by parallelizing the
code on the Beowulf cluster with the Message Passing Interface (MPI)~\cite%
{MPI}. The simplest way of parallelizing a QMC code is to make smaller
number of measurements on each node and to average the results of each node
to obtain the final result effectively with the desired number of
measurements. In the CDMFT+QMC algorithm, this means that the heavy exchange
of information between processors occurs at step (5) above. In Fig.~\ref%
{MPI_SPEEDUP.fig} the combined total number of measurements is 64,000 for
the circles, which means 2,000 measurements on each node for calculations
with 32 nodes. The switch is at 10 Gb/s on infiniband and the processors are
3.6 GHz dual core Xeon. For a small number of nodes the speedup appears
nearly perfect. As the number of nodes increases, it starts to deviate from
the perfect line because the unparallelized part of the code starts to
compensate the speedup. Speedup with less number of measurements on each
node (diamonds) deviates further from the dashed line. Most
of the calculations in the present work were done with 16 or 32 nodes and
with up to 128,000 measurements.

\section{Comparison with Exact Results (Benchmarking)}

\label{section5}

\label{section5-1}

The one-dimensional Hubbard model represents an ideal benchmark for the
current and other cluster methods for several reasons. First, there exist
several (analytically and numerically) exact results to compare with.
Second, as mentioned before, the CDMFT scheme is expected to be in the worst
case scenario in one-dimension, so that if it reproduces those exact results
it is likely to capture the physics more accurately in higher dimensions.
Third, a study of a systematic size dependence is much easier because of the
linear geometry. 
\begin{figure}[tbp]
\includegraphics[width=8.0cm]{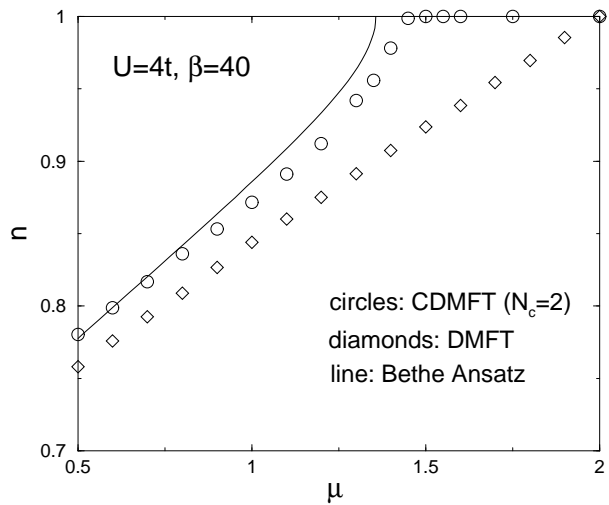}
\caption{Density $n$ as a function of the chemical potential $\protect\mu $
in the one-dimensional Hubbard model for $U/t=4$, $\protect\beta =40$, $%
N_{c}=2$ (circles). The diamonds are obtained within the single site DMFT
with the same parameters, while the solid curve is computed by the Bethe
ansatz at zero temperature. }
\label{mu_n_1D_QMC_beta=40.fig}
\end{figure}

In Fig.~\ref{mu_n_1D_QMC_beta=40.fig} we show the density $n$ as a function
of the chemical potential $\mu $ in the one-dimensional Hubbard model for $%
U/t=4$. It shows that CDMFT on the smallest cluster ($N_{c}=2$) already
captures with high accuracy the evolution of the density as a function of
the chemical potential and the compressibility divergence at the Mott
transition~\cite{Comment:1D}, in good agreement with the exact Bethe
ansatz result (solid curve)~\cite{LW:1968}. This feature is apparently
missed in the single site DMFT (diamonds) which also misses the Mott gap at
half-filling for $U/t=4$. The deviation from the exact location where the
density suddenly drops seems to be caused by a finite-size effect since we
have checked that it does not come from finite temperature or from the
imaginary-time discretization. The compressibility divergence as well as the
Mott gap at half-filling for $U/t=4$ were recently reproduced by Capone 
\textit{et al.}~\cite{CCKCK:2004} using ED technique at zero temperature. 
\begin{figure}[tbp]
\includegraphics[width=8.0cm]{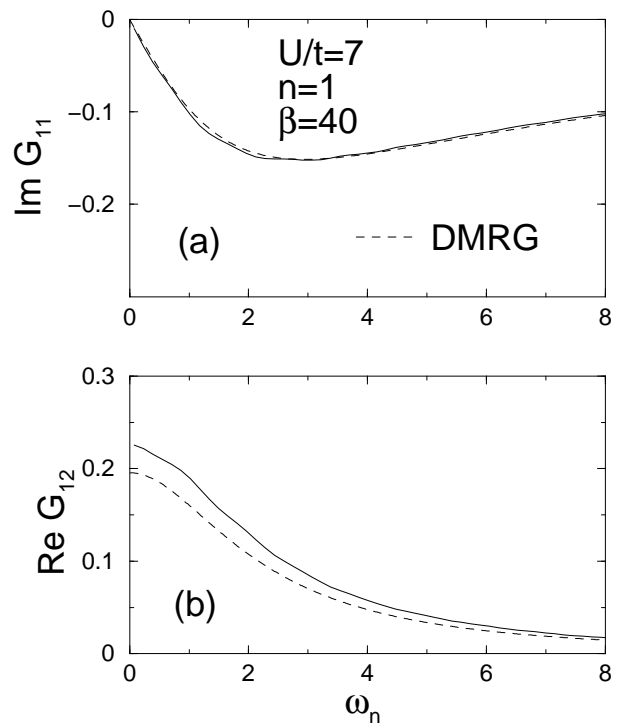}
\caption{(a) Imaginary part of the local Green's function $G_{11}$ and (b)
real part of the nearest neighbor Green's function $G_{12}$ in the
one-dimensional Hubbard model for $U/t=7$, $n=1$ and $\protect\beta =40$ on $%
N_{c}=2$ cluster. The dashed curves are DMRG results. }
\label{G11_G12_U=7_beta=40.fig}
\end{figure}

In Fig.~\ref{G11_G12_U=7_beta=40.fig} the imaginary part of the local
Green's function $G_{11}$ and the real part of the nearest-neighbor Green
function $G_{12}$ are compared on the Matsubara axis with DMRG results shown
as dashed curves. CDMFT with $N_{c}=2$ closely follows the DMRG on the whole
Matsubara axis, and the two results become even closer for $N_{c}=4$ (not
shown here). These results present an independent confirmation of the
ability of CDMFT to reproduce the exact results in one-dimension with small
clusters. This is very encouraging, since mean field methods are expected to
perform even better as the dimensionality increases. In the application
section on the Luttinger liquid, we will study quantities that are more
sensitive to the size dependence.

\section{Convergence with system size}

\label{section_convergence}

\subsection{One-dimensional Hubbard Model}

\label{section5-2} 
\begin{figure}[tbp]
\includegraphics[width=8.0cm]{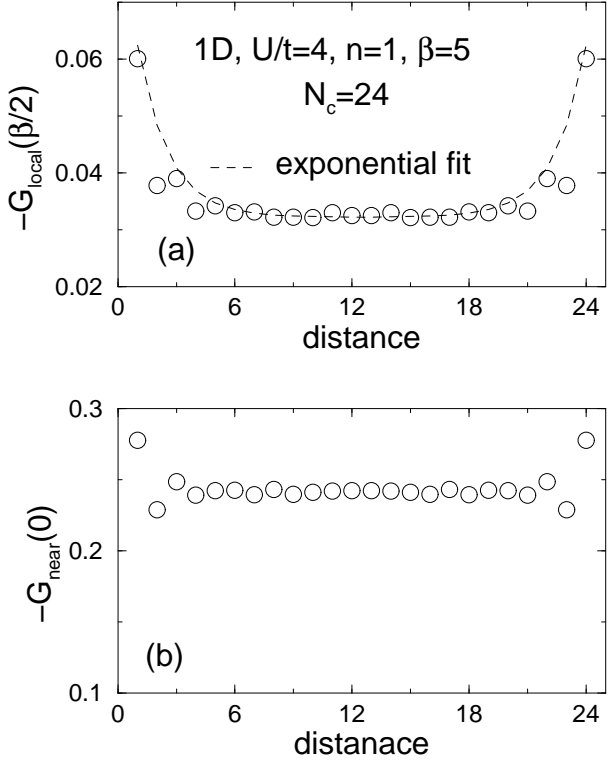}
\caption{(a) Local Green's function $G_{local}(\protect\tau =\protect\beta %
/2)$ and (b) nearest neighbor Green's function $G_{near}(\protect\tau =0)$
as a function of distance from the boundary of a linear chain with $N_{c}=24$ in
the one-dimensional Hubbard model for $U/t=4$, $n=1$, $\protect\beta =5$
(circles). The dashed curve is the exponential fit. }
\label{G_tau_1D_U=4_n=1_beta=5_nc=24.fig}
\end{figure}
Fig.~\ref{G_tau_1D_U=4_n=1_beta=5_nc=24.fig} shows the local Green function $%
G_{local}(\tau =\beta /2)$ and the nearest-neighbor Green function $%
G_{near}(\tau =0)$ as a function of distance from the boundary of a long linear
chain ($N_{c}=24$). The local and nearest-neighbor Green functions rapidly
(exponentially) approach the infinite cluster limit a few lattice sites away
from the boundary. The largest deviation from the
infinite cluster limit occurs essentially at the boundary. This feature has
lead to recent attempts~\cite{PK:2005} to greatly improve the convergence
properties of CDMFT at large clusters by weighting more near the center of
the cluster. It is called weighted-CDMFT, an approach which is being
developed at present~\cite{PK:2005}. In this paper we focus on small
clusters and calculate lattice quantities without weighting (Eq.~\ref{eq40})
and study how much correlation effect is captured by small clusters compared
with the infinite size cluster. Because most of the detailed study of the
Hubbard model~\cite{MJPH:2004,CCKPK:2005,KKSTCK:2005} in the physically
relevant regime (intermediate to strong coupling and low temperature) have
been obtained only on small clusters, the present study will show how much
those results represent the infinite cluster limit.

\begin{figure}[tbp]
\includegraphics[width=8.0cm]{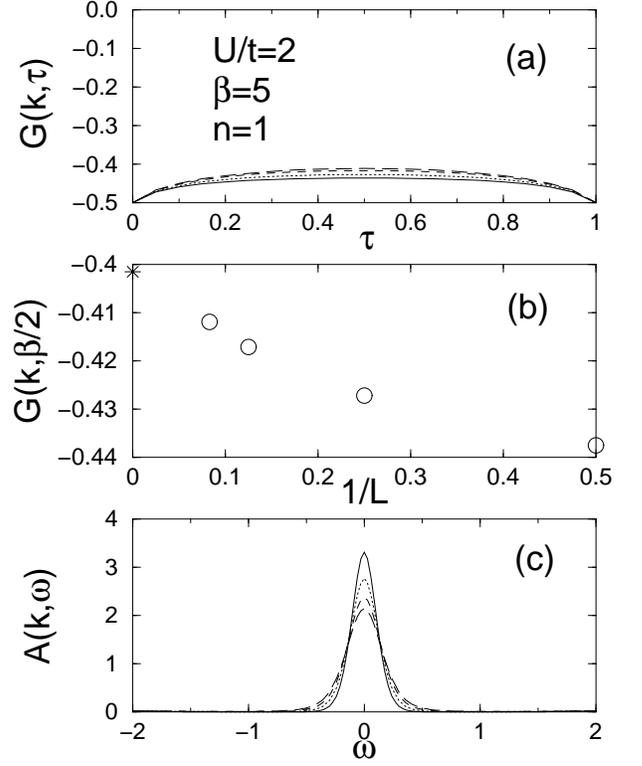}
\caption{(a) Imaginary-time Green's function $G(\vec{k},\protect\tau )$ at
the Fermi point ($\vec{k}=\protect\pi /2$) in the one-dimensional Hubbard
model for $U/t=2$, $\protect\beta =5$, $n=1$ with $N_{c}=2,4,8,12$ (solid,
dotted, dashed, long-dashed curves). (b) Cluster size ($N_{c}=L$) dependence
of $G(\vec{k},\protect\beta /2)$ at small clusters. (c) The corresponding
spectral function $A(\vec{k},\protect\omega )$. The star in (b) represents
the infinite cluster limit extracted from large clusters. }
\label{G_tau_w_1D_U=2_n=1_beta=5.fig}
\end{figure}

Figure~\ref{G_tau_w_1D_U=2_n=1_beta=5.fig}(a) shows the imaginary-time Green
function $G(\vec{k},\tau )$ at the Fermi point ($\vec{k}=\pi /2$) for $U/t=2$%
, $\beta =5$, $n=1$ with $N_{c}=2,4,8,12$. As the cluster size increases, $G(%
\vec{k},\tau )$ becomes smaller in magnitude and the infinite cluster limit is
approached in a way opposite to that in finite size simulations, a
phenomenon that was observed before in the Dynamical Cluster Approximation
(DCA)~\cite{MHJ:2000}. 

The quantity $G(\vec{k},\beta /2)$ at the Fermi wave vector is a useful
measure of the strength of correlations. It varies from $-1/2$ for $U=0$ to $%
0$ for $U=\infty $ at half-filling. Thus throughout the paper $\left( U\neq
0\right) $ the \textquotedblleft correlation ratio\textquotedblright\ 
\begin{equation}
C_{r}\equiv \left[ \frac{G(\vec{k},\beta /2)|_{N_{c}}+1/2}{G(\vec{k},\beta
/2)|_{N_{c}=\infty }+1/2}\right]   \label{eq170}
\end{equation}%
will be used as an approximate estimate of how much the correlation effects
are captured by a given cluster of size $N_{c}$, compared with the infinite
cluster. $C_{r}$ is equal to unity when finite-size effects are absent.

Figure~\ref{G_tau_w_1D_U=2_n=1_beta=5.fig}(b) shows the cluster size ($%
N_{c}=L$) dependence of $G(\vec{k},\beta /2)$ for small clusters. At small $%
L $ the curvature is upward so that $G(\vec{k},\beta /2)$ is much closer to
the value of the infinite size cluster than what would be naively
extrapolated from large clusters. The corresponding spectral function $A(%
\vec{k},\omega )$ in Fig.~\ref{G_tau_w_1D_U=2_n=1_beta=5.fig}(c) shows a
peak at the Fermi level for all clusters up to $L=12$. Although this looks
like a quasiparticle peak, it is disproved by a close inspection of the
corresponding self-energy.

We extract the self-energy from the lattice spectral function $A(\vec{k}%
,\omega )$, and the relation between $A(\vec{k},\omega )$ and $G(\vec{k}%
,\omega )$ 
\begin{eqnarray}
G(\vec{k},\omega ) &=&\int d\omega ^{\prime }\frac{A(\vec{k},\omega ^{\prime
})}{\omega +i\delta -\omega ^{\prime }}\;,  \notag \\
G(\vec{k},\omega ) &=&\frac{1}{\omega +i\delta -\varepsilon _{\vec{k}%
}-\Sigma (\vec{k},\omega )} \;, \label{eq180}
\end{eqnarray}%
where $\delta $ is an infinitesimally small positive number and $\varepsilon
_{\vec{k}}$ is the non-interacting energy dispersion. 
\begin{figure}[tbp]
\includegraphics[width=8.0cm]{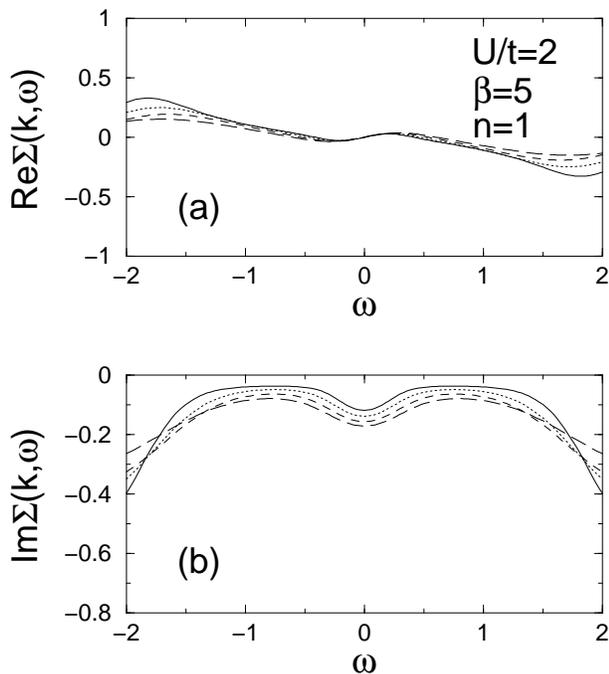}
\caption{Real (a) and imaginary (b) part of the self-energy $\Sigma (\vec{k},%
\protect\omega )$ at the Fermi point ($\vec{k}=\protect\pi /2$) in the
one-dimensional Hubbard model for $U/t=2$, $\protect\beta =5$, $n=1$ with $%
N_{c}=2,4,8,12$ (solid, dotted, dashed, long-dashed curves). }
\label{S_w_1D_U=2_n=1_beta=5.fig}
\end{figure}
In spite of the peak in $A(\vec{k},\omega )$, the corresponding self-energy
in Fig.~\ref{S_w_1D_U=2_n=1_beta=5.fig} shows not only a local maximum at $%
\omega =0$ in the absolute value of the imaginary part, but also a positive
slope at the Fermi level in the real part, which cannot be reconciled with a
Fermi liquid. The scattering rate increases with increasing cluster size, as
found in DCA~\cite{MHJ:2000}, in contrast to the results of finite size
simulations.

\begin{figure}[tbp]
\includegraphics[width=8.0cm]{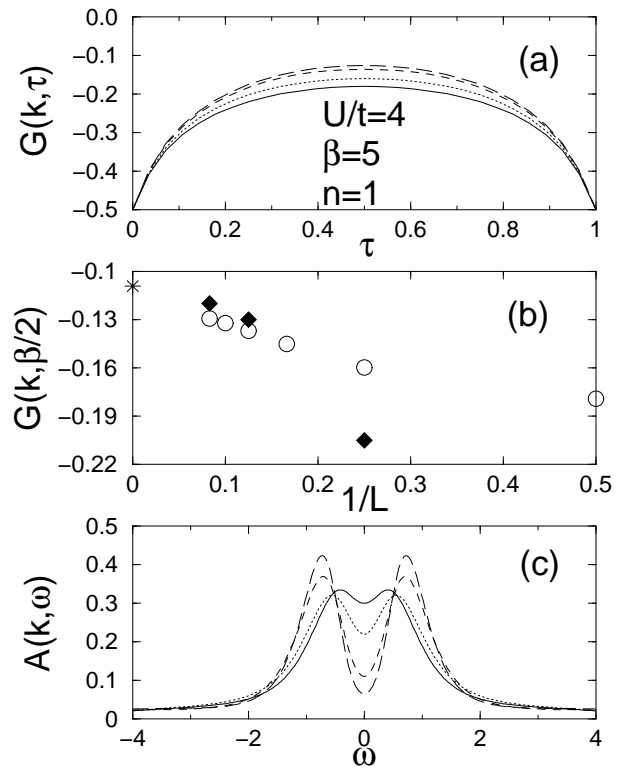}
\caption{(a) Imaginary-time Green's function $G(\vec{k},\protect\tau )$ at
the Fermi point ($\vec{k}=\protect\pi /2$) in the one-dimensional Hubbard
model for $U/t=4$, $\protect\beta =5$, $n=1$ with $N_{c}=2,4,8,12$ (solid,
dotted, dashed, long-dashed curves). (b) Cluster size ($N_{c}=L$) dependence
of $G(\vec{k},\protect\beta /2)$ at small clusters. The filled diamonds are
DCA results in Ref.~\protect\cite{MHJ:2000} with the same parameters. (c)
The corresponding spectral function $A(\vec{k},\protect\omega )$. The star
in (b) represents the infinite cluster limit extracted from large clusters. }
\label{G_tau_w_1D_U=4_n=1_beta=5.fig}
\end{figure}

\begin{figure}[tbp]
\includegraphics[width=8.0cm]{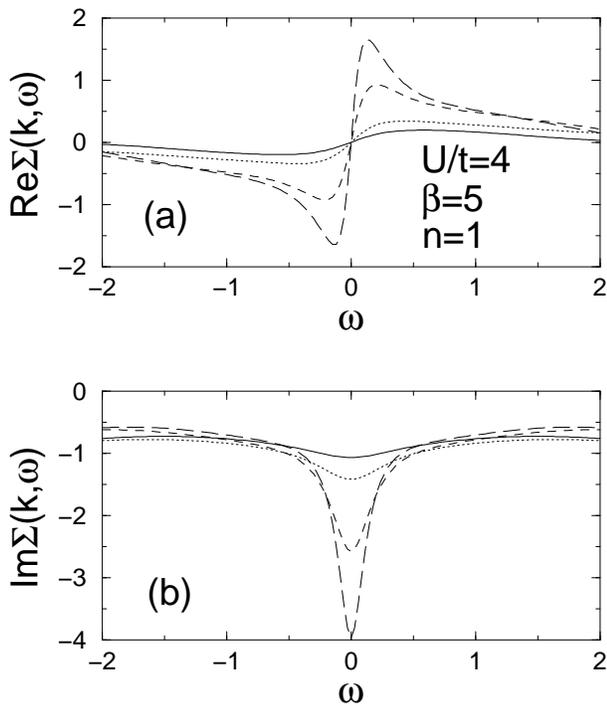}
\caption{Real (a) and imaginary (b) part of the self-energy $\Sigma (\vec{k},%
\protect\omega )$ at the Fermi point ($\vec{k}=\protect\pi /2$) in the
one-dimensional Hubbard model for $U/t=4$, $\protect\beta =5$, $n=1$ with $%
N_{c}=2,4,8,12$ (solid, dotted, dashed, long-dashed curves). }
\label{S_w_1D_U=4_n=1_beta=5.fig}
\end{figure}

For the more correlated case of $U/t=4$ ($U$ equal to the bandwidth in
one-dimension) in Fig.~\ref{G_tau_w_1D_U=4_n=1_beta=5.fig}, $G(\vec{k},\tau
) $ becomes much smaller in magnitude than $1/2$, the result for an
uncorrelated system. A similar cluster size dependence is also found here.
In Fig.~\ref{G_tau_w_1D_U=4_n=1_beta=5.fig}(b) we compare our cluster size
dependence with that of DCA~\cite{MHJ:2000} (filled diamonds) for small
clusters. Generally the curvatures are opposite, namely, upward in CDMFT and
downward in DCA. This upward curvature enables even an $L=2$ cluster to
capture, as measured by $C_{r}$, about $82\%$ of the correlation effect of
the infinite size cluster. The corresponding spectral function $A(\vec{k}%
,\omega )$ already shows a pseudogap for $L=2$, in contrast to the DCA
result~\cite{MHJ:2000} in which a pseudogap begins to appear for $L=8$. 
For $U/t=4$ the scattering rate in Fig.~\ref{S_w_1D_U=4_n=1_beta=5.fig} is
large enough to create the pseudogap in $A(\vec{k},\omega )$ for all
clusters. 

\begin{figure}[tbp]
\includegraphics[width=8.0cm]{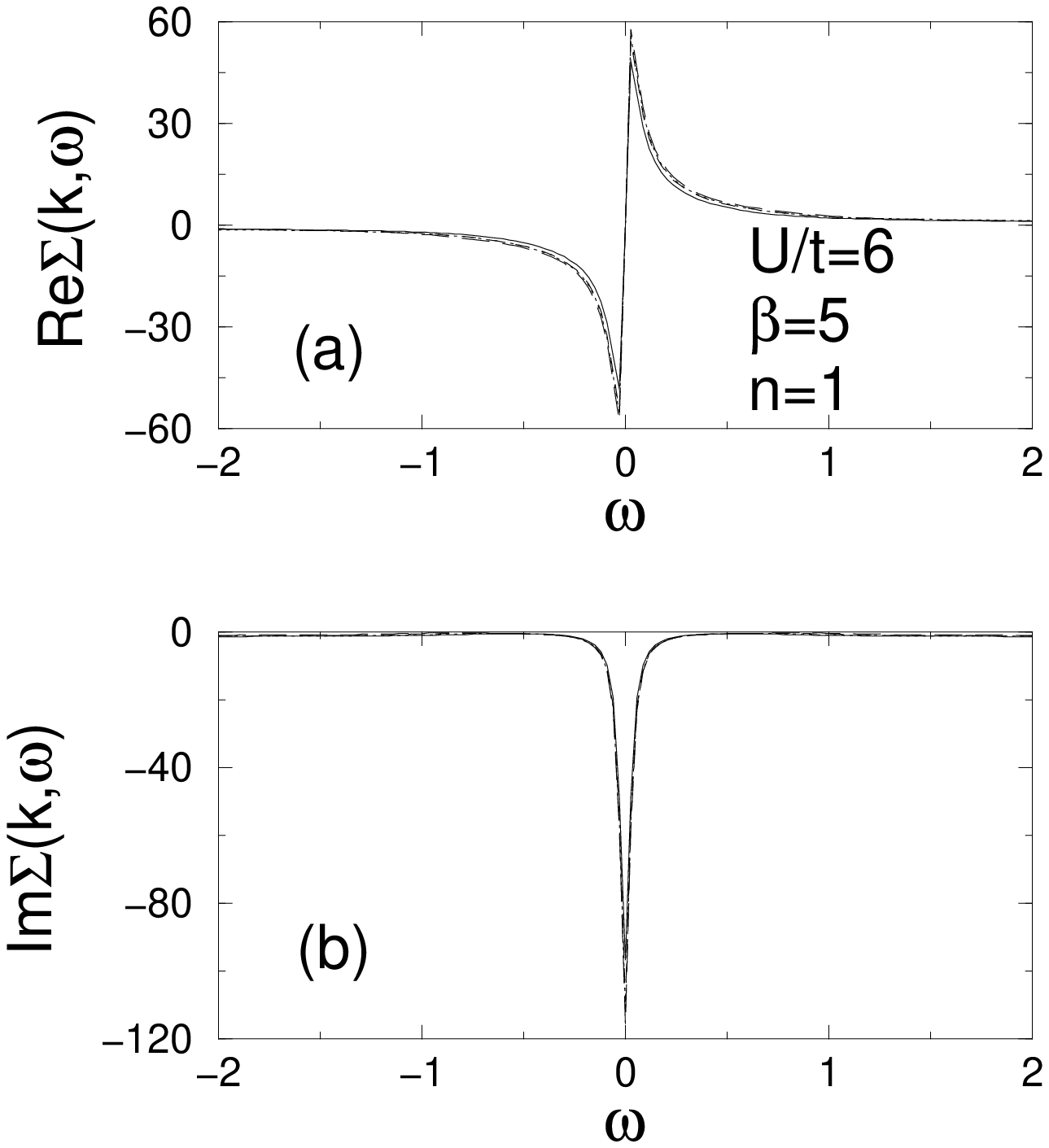}
\caption{Real (a) and imaginary (b) part of the self-energy $\Sigma (\vec{k},%
\protect\omega )$ at the Fermi point ($\vec{k}=\protect\pi /2$) in the
one-dimensional Hubbard model for $U/t=6$, $\protect\beta =5$, $n=1$ with $%
N_{c}=2,4,8,12$ (solid, dotted, dashed, long-dashed curves). }
\label{S_w_1D_U=6_n=1_beta=5.fig}
\end{figure}

For an even more correlated case of $U/t=6$ in Fig.~\ref%
{G_tau_w_1D_U=6_n=1_beta=5.fig}, an $L=2$ cluster captures $99\%$ of the
correlation effect (as measured by Eq.~\ref{eq170}) of the infinite size
cluster. Thus, at intermediate to strong coupling, short-range correlation
effect (on a small cluster) starts to dominate the physics in the single particle spectral
function, reinforcing our recent results~\cite{KKSTCK:2005} based on the
two-dimensional Hubbard model. The corresponding $A(\vec{k},\omega )$ shows
a large pseudogap (or real gap) for all cluster sizes. 
\begin{figure}[tbp]
\includegraphics[width=8.0cm]{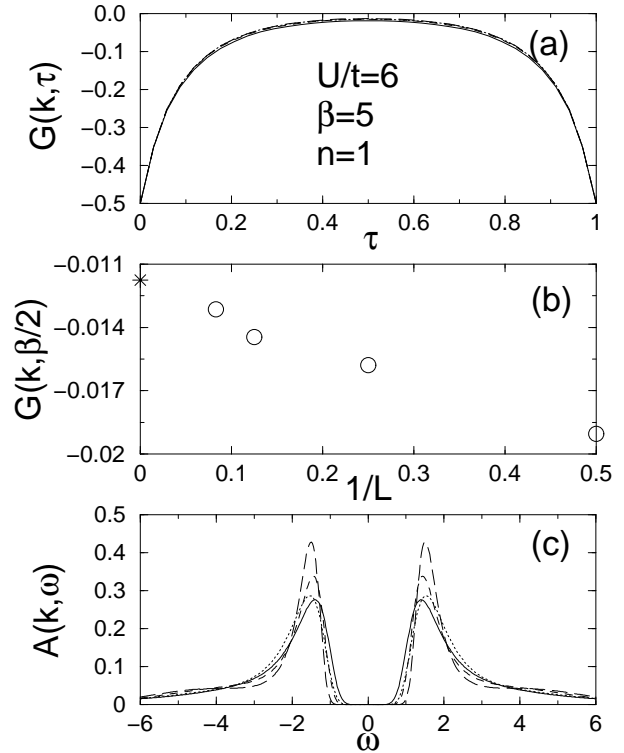}
\caption{(a) Imaginary-time Green's function $G(\vec{k},\protect\tau )$ at
the Fermi point ($\vec{k}=\protect\pi /2$) in the one-dimensional Hubbard
model for $U/t=6$, $\protect\beta =5$, $n=1$ with $N_{c}=2,4,8,12$ (solid,
dotted, dashed, long-dashed curves). (b) Cluster size ($N_{c}=L$) dependence
of $G(\vec{k},\protect\beta /2)$ for small clusters. (c) The corresponding
spectral function $A(\vec{k},\protect\omega )$. The star in (b) represents
the infinite cluster limit extracted from large clusters. }
\label{G_tau_w_1D_U=6_n=1_beta=5.fig}
\end{figure}
The huge scattering rate in Fig.~\ref{S_w_1D_U=6_n=1_beta=5.fig} at the
Fermi energy is responsible for the large pseudogap (or real gap) in the spectral
function. 

\begin{figure}[tbp]
\includegraphics[width=8.0cm]{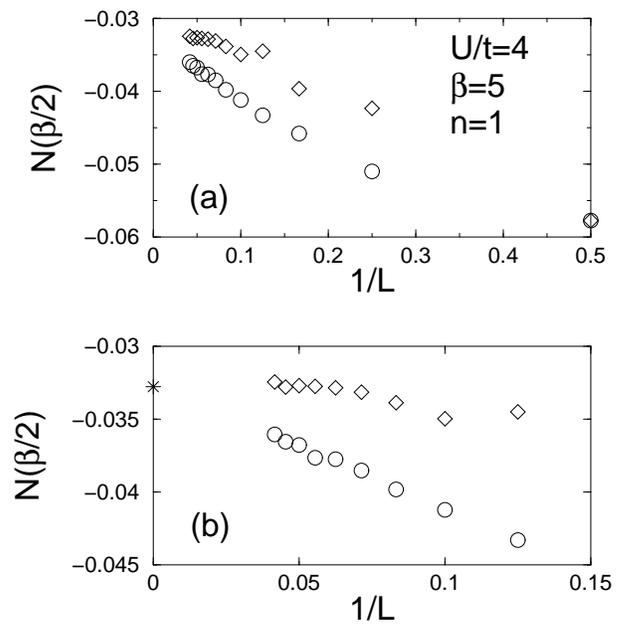}
\caption{ (a) Cluster size ($N_{c}=L$) dependence of the imaginary-time
density of states $N(\protect\tau )$ at $\protect\tau =\protect\beta /2$ in
the one-dimensional Hubbard model for $U/t=4$, $\protect\beta =5$, $n=1$.
The circles are obtained from the average of the lattice Green's function
(without weighting) over the Brillouin zone, while the diamonds are
calculated from the local Green's function at the center of the cluster (a
linear chain in one-dimension). (b) Close up of the region at large $L$. The
star in (b) represents the infinite cluster limit extracted by a linear
extrapolation at large clusters. }
\label{N_tau_1D_U=4_n=1_beta=5.fig}
\end{figure}

Recently there has been a debate about the convergence of the two quantum
cluster methods (CDMFT and DCA)~\cite{BK:2002,MJ:2002,AMJ:2005} using a
highly simplified one-dimensional large-N model Hamiltonian where dynamics
are completely suppressed in the limit of $N\rightarrow \infty $. The
general consensus about the convergence of CDMFT (based on the study of this
model Hamiltonian) is that purely local quantities defined on central
cluster sites converge exponentially, while lattice quantities such as the
lattice Green function converge with corrections of order $1/L$. Here we
address this issue with a more realistic Hamiltonian, the one-dimensional
Hubbard model at intermediate coupling of $U/t=4$. Figure~\ref%
{N_tau_1D_U=4_n=1_beta=5.fig}(a) shows the cluster size dependence of the
imaginary-time density of states $N(\tau )$ at $\tau =\beta /2$. We obtained 
$N(\tau )$ in two different ways: Taking the average of the lattice Green
function (obtained without weighting) over the Brillouin zone (circles) and
taking the local Green function at the center of the cluster (diamonds). For 
$N_{c}=2$ they are identical while for larger clusters, $N(\beta /2)$
obtained from the local Green's function approaches the $N_{c}=\infty $
limit much faster than that from the lattice Green function that converges
linearly in $1/L$~\cite{Comment:Weighting}, in agreement with the previous
results based on the large-N model. In spite of this slow convergence, the
slope is so small that the $N_{c}=2$ cluster already accounts for $95\%$ of
the correlation effect of the infinite cluster (using $C_{r}$ as a measure
with $G(\vec{k},\beta /2)\rightarrow N(\beta /2)$), much larger than $82\%$
for $G(\vec{k},\beta /2)$. Figure~\ref{N_tau_1D_U=4_n=1_beta=5.fig}(b) is a
close up of Fig.~\ref{N_tau_1D_U=4_n=1_beta=5.fig}(a) at large $L$. $N(\beta
/2)$ from the two methods converge to a single value as $L\rightarrow \infty 
$. $N(\beta /2)$ from the local Green function approaches the infinite-size
limit much faster than $1/L^{2}$ and apparently converges exponentially. 


\subsection{Two-dimensional Hubbard Model}

\label{section5-3}

\begin{figure}[tbp]
\includegraphics[width=8.0cm]{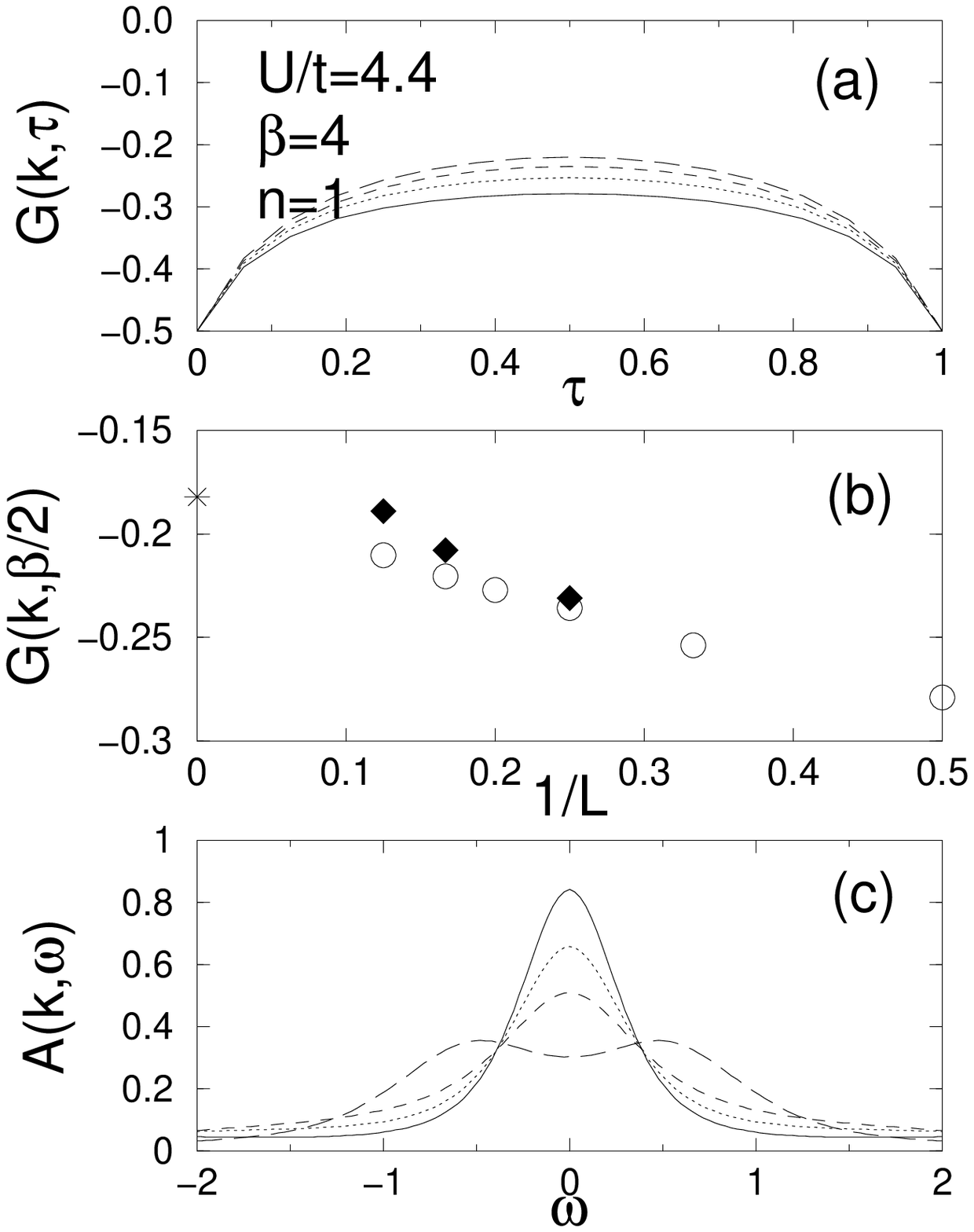}
\caption{(a) Imaginary-time Green's function $G(\vec{k},\protect\tau )$ at
the Fermi surface ($\vec{k}=(\protect\pi ,0)$) in the two-dimensional
Hubbard model for $U/t=4.4$, $\protect\beta =4$, $n=1$ with $N_{c}=2\times
2,3\times 3,4\times 4,6\times 6$ (solid, dotted, dashed, long-dashed
curves). $\Delta \protect\tau =0.25$ is used here. (b) Cluster size ($%
N_{c}=L\times L$) dependence of $G(\vec{k},\protect\beta /2)$ at small
clusters. 
The filled diamonds are DCA results of Ref.\protect\cite{JMHM:2001}. 
(c) The corresponding spectral function $A(\vec{k},\protect\omega )
$. The star in (b) represents the infinite cluster limit extracted by a linear
extrapolation at large clusters. }
\label{G_tau_w_2D_U=4.4_n=1_beta=4.fig}
\end{figure}

The two-dimensional Hubbard Hamiltonian on a square lattice has been
intensively studied for many years, especially since Anderson's seminal
paper~\cite{Anderson:1987} on high temperature superconductivity. There is
mounting evidence that this model correctly describes the low-energy physics
of the copper oxides~\cite{TKS:2005}. In addition to various types of
long-range order observed in the cuprates, one must understand the
intriguing normal state pseudogap~\cite{TS:1999} in the underdoped regime.
In this section we focus on the size dependence of the spectral function for
the half-filled two-dimensional Hubbard model for small clusters at finite
temperature. Figure~\ref{G_tau_w_2D_U=4.4_n=1_beta=4.fig}(a) shows the
imaginary-time Green's function $G(\vec{k},\tau )$ at the Fermi surface ($%
\vec{k}=(\pi ,0)$) for $U/t=4.4$, $\beta =4$, $n=1$ with $N_{c}=2\times
2,3\times 3,4\times 4,6\times 6$~\cite{Comment:2D}. As the cluster size
increases, $G(\vec{k},\tau )$ decreases in magnitude, as in the
one-dimensional case, a behavior opposite to that of finite size
simulations. This trend is in agreement with DCA~\cite{JMHM:2001}, as shown
in Figure~\ref{G_tau_w_2D_U=4.4_n=1_beta=4.fig}(b). The spectral weight $A(%
\vec{k},\omega )$ for the same parameters shows a peak at $\omega =0$ for
small $L$ ($N_{c}=L\times L$) but starts exhibiting a pseudogap for $L\geq 6$%
. This is consistent with our recent results with CDMFT+ED~\cite{KKSTCK:2005}
where we find that at weak coupling a large correlation length (on a large
cluster) is required to create a pseudogap. From the Two-Particle
Self-Consistent (TPSC) approach, we know that to obtain a pseudogap in this
regime of coupling strength, the antiferromagnetic correlation length has to
be larger than the single-particle thermal de Broglie wave length~\cite%
{Vilk:1997, Vilk:1996}. 
Unlike in the one-dimensional case, the imaginary part of the self-energy
for $L\leq 4$ has a very shallow maximum or a minimum at the Fermi level,
accompanied by a negative slope in the real part as seen in Fig.~\ref%
{S_w_2D_U=4.4_n=1_beta=4.fig}. This feature is consistent with a Fermi
liquid at finite temperature. For $L\geq 6$, however, the scattering rate
has a local maximum together with a large positive slope in the real part,
resulting in the pseudogap in the spectral function. 
\begin{figure}[tbp]
\includegraphics[width=8.0cm]{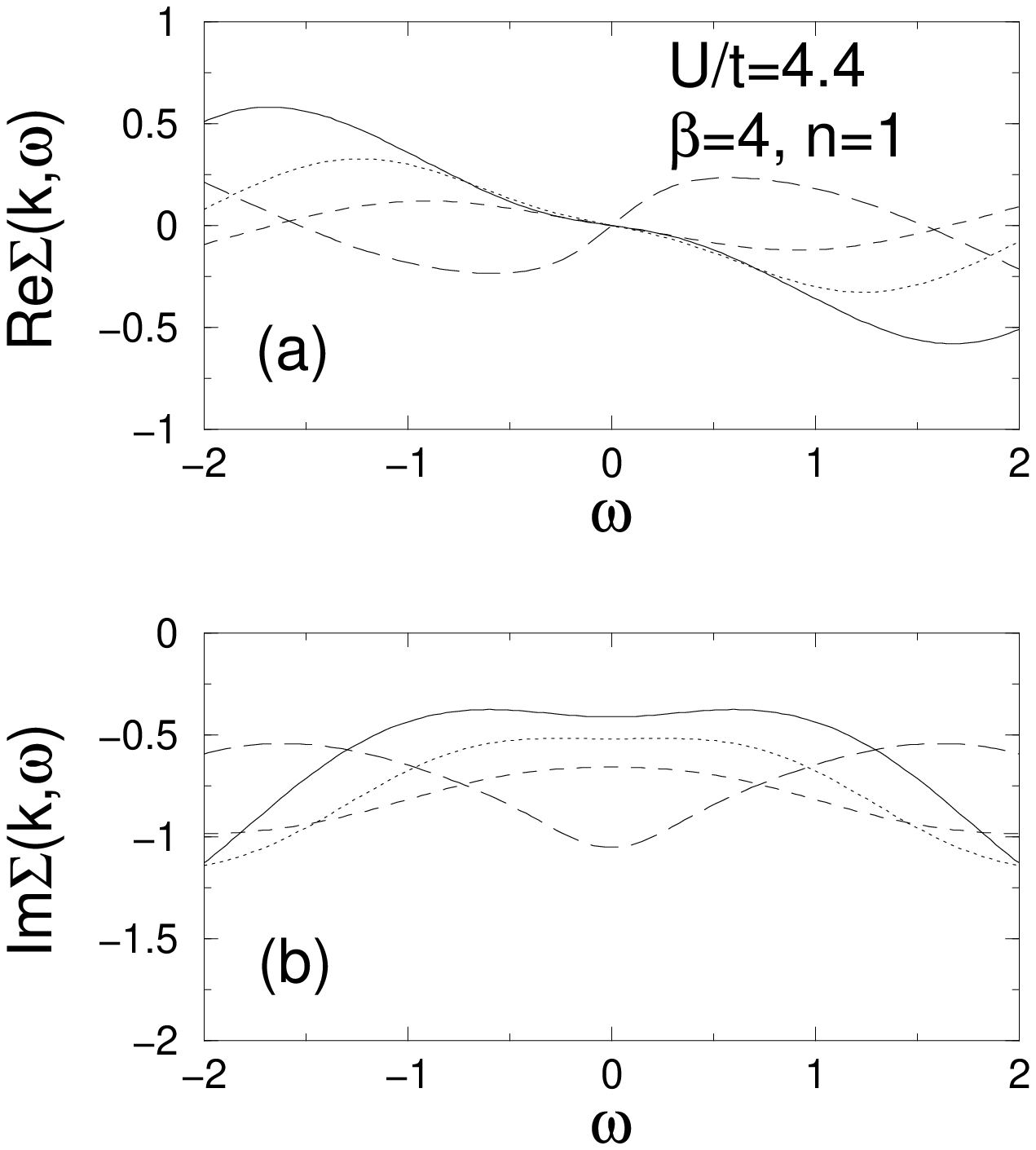}
\caption{Real (a) and imaginary (b) part of the self-energy $\Sigma (\vec{k},%
\protect\omega )$ at the Fermi surface ($\vec{k}=(\protect\pi ,0)$) in the
two-dimensional Hubbard model for $U/t=4.4$, $\protect\beta =4$, $n=1$ with $%
N_{c}=2\times 2,3\times 3,4\times 4,6\times 6$ (solid, dotted, dashed,
long-dashed curves). }
\label{S_w_2D_U=4.4_n=1_beta=4.fig}
\end{figure}

\begin{figure}[tbp]
\includegraphics[width=8.0cm]{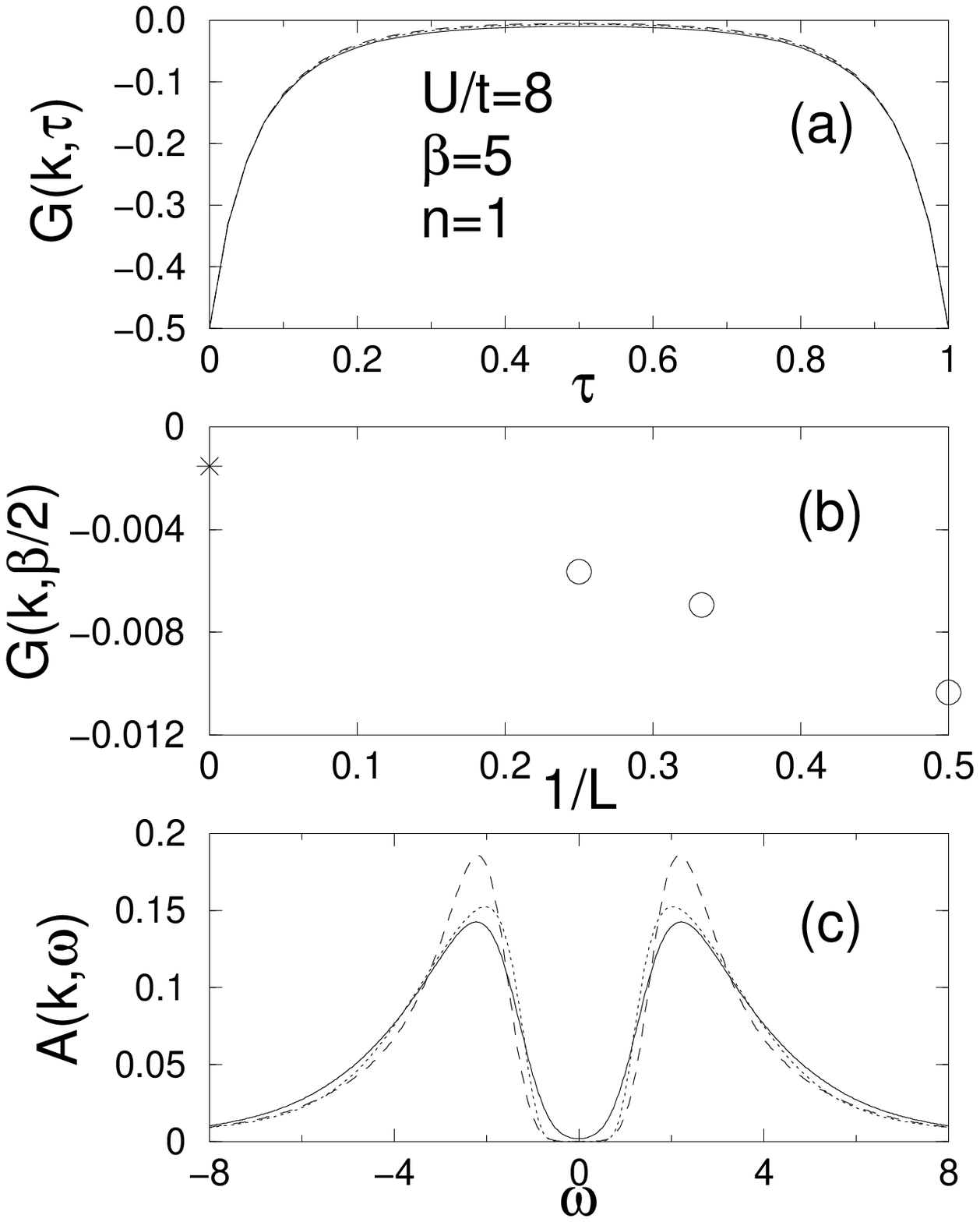}
\caption{(a) Imaginary-time Green's function $G(\vec{k},\protect\tau )$ at
the Fermi surface ($\vec{k}=(\protect\pi ,0)$) in the two-dimensional
Hubbard model for $U/t=8$, $\protect\beta =5$, $n=1$ with $N_{c}=2\times
2,3\times 3,4\times 4$ (solid, dotted, dashed curves). (b) Cluster size ($%
N_{c}=L\times L$) dependence of $G(\vec{k},\protect\beta /2)$ for small
clusters. (c)
The corresponding spectral function $A(\vec{k},\protect\omega )$. The star
in (b) represents the infinite cluster limit extracted by a linear
extrapolation. }
\label{G_tau_w_2D_U=8_n=1_beta=5.fig}
\end{figure}

\begin{figure}[tbp]
\includegraphics[width=8.0cm]{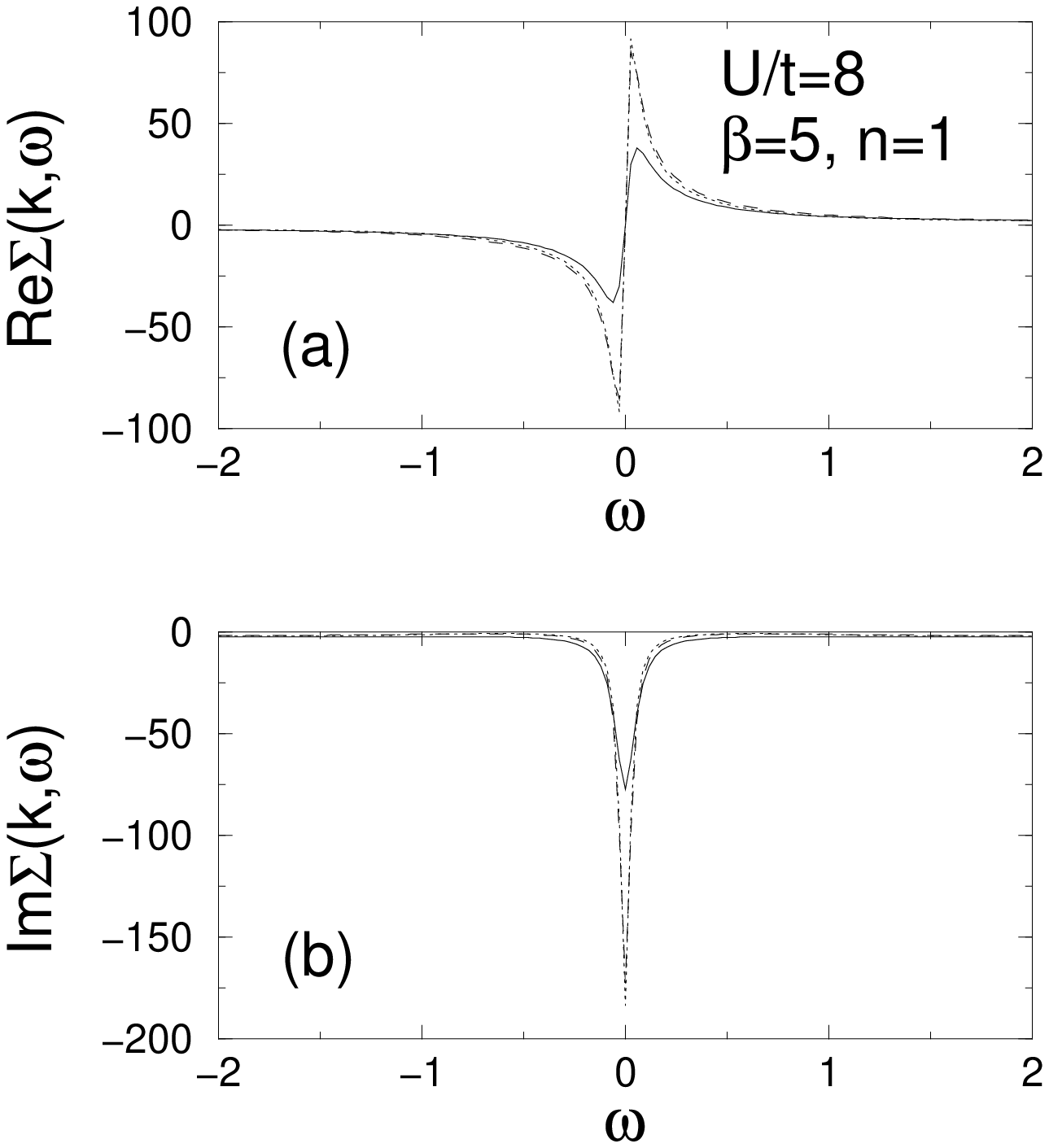}
\caption{Real (a) and imaginary (b) part of the self-energy $\Sigma (\vec{k},%
\protect\omega )$ at the Fermi surface ($\vec{k}=(\protect\pi ,0)$) for the
two-dimensional Hubbard model with $U/t=8$, $\protect\beta =5$, $n=1$ for $%
N_{c}=2\times 2,3\times 3,4\times 4$ (solid, dotted, dashed curves). }
\label{S_w_2D_U=8_n=1_beta=5.fig}
\end{figure}

Next we study the more correlated case of $U/t=8$ in Fig.~\ref%
{G_tau_w_2D_U=8_n=1_beta=5.fig}. This regime is believed to be relevant for
the hole-doped cuprates. When $U$ becomes equal to the bandwidth, the
cluster size dependence of $G(\vec{k},\tau )$ is extremely weak. As can be
seen in Fig.~\ref{G_tau_w_2D_U=8_n=1_beta=5.fig}(b), $N_{c}=2\times 2$
already accounts for more than $95\%$ of the correlation effect (as measured
by Eq.~\ref{eq170}) of the infinite size cluster in the single particle
spectrum, supporting our recent result obtained with CDMFT+ED method~\cite%
{KKSTCK:2005} in the two-dimensional Hubbard model. The large gap in $A(\vec{%
k},\omega )$ does not change significantly with increasing cluster size. 
The self-energy for $U/t=8$ (Fig.~\ref{S_w_2D_U=8_n=1_beta=5.fig}) appears similar to what was found in the
one-dimensional case with $U/t=6$ where the imaginary part has a very large
peak at the Fermi energy, leading to what appears as a large gap in $A(\vec{k%
},\omega )$. 
When short-range spatial correlations are treated explicitly, the
well-known metal-insulator transition in the single site DMFT disappears
immediately as shown in our recent articles~\cite{KKSTCK:2005,KGT:2005}.
Frustration would restore the metal-insulator
transition~\cite{PBK:2004}.


\section{Spinons and holons in CDMFT}

\label{section5-4}

Spinon and holon dispersions were recently found experimentally in a
quasi-one-dimensional organic conductors away from half-filling by Claessen 
\textit{et al.}~\cite{CSSBDJ:2002}. These separate features of the
dispersion in a Luttinger liquid are a challenge for numerical approaches.
Indeed, we know from bosonization and from the renormalization group~\cite%
{Voit:1995} that they arise from long-wavelength physics, hence it is is not
obvious how these features can come out from small cluster calculations.
They have been seen theoretically in the one-dimensional Hubbard model away
from half-filling by Benthien \textit{et al.}~\cite{BGJ:2004} using
Density-Matrix Renormalization Group. The evidence from straight QMC
calculations is based on the analysis of chains of size 64~\cite{ZAHS:1998}.
On the other hand, with Cluster Perturbation Theory one finds clear signs of
the holon and spinon dispersion at zero temperature already for clusters of
size 12~\cite{SPP:2000}.

\begin{figure}[tbp]
\resizebox{7.8cm}{!}{\includegraphics{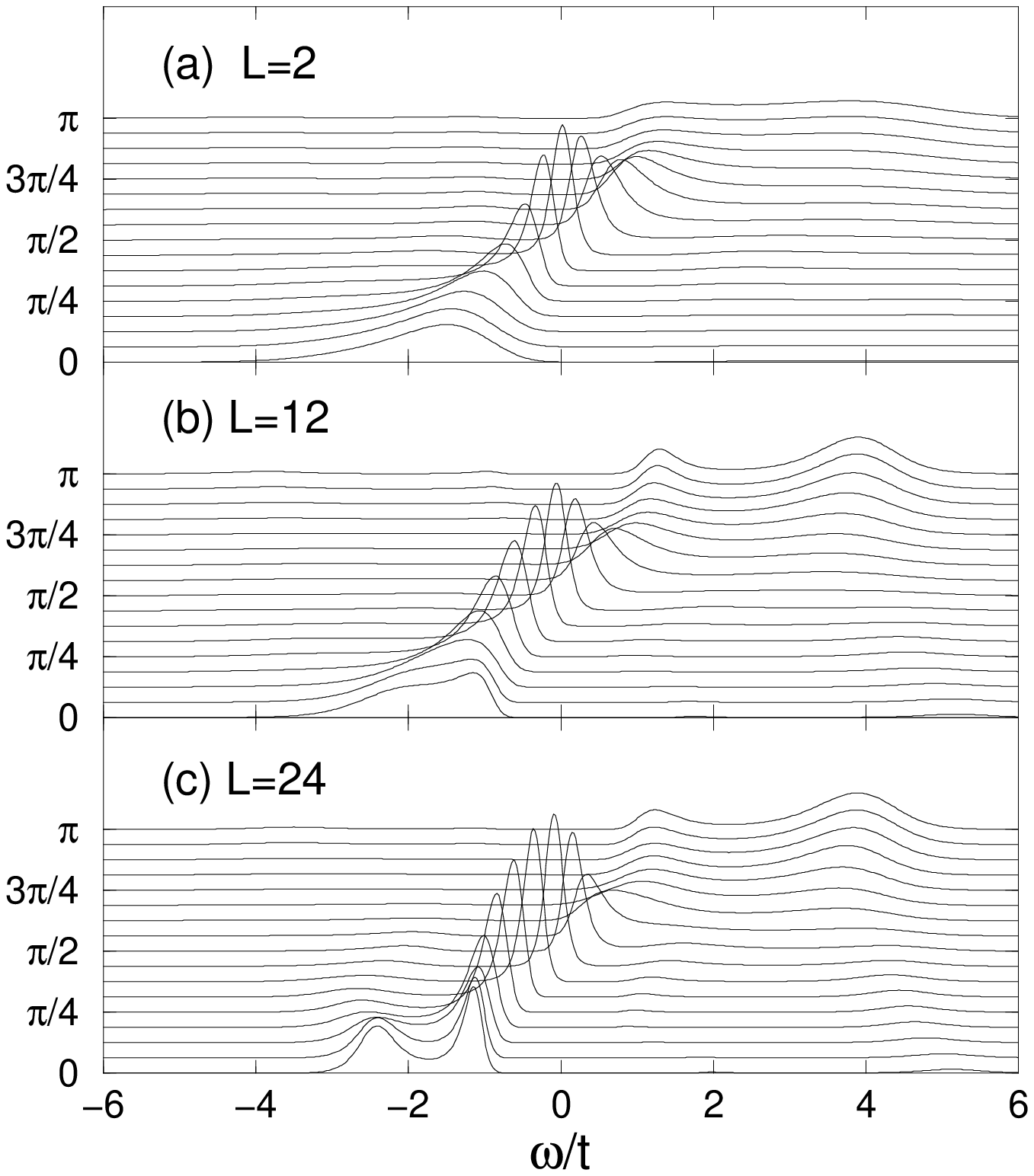}}
\par\hspace{0.2cm}
\includegraphics[width=8.0cm]{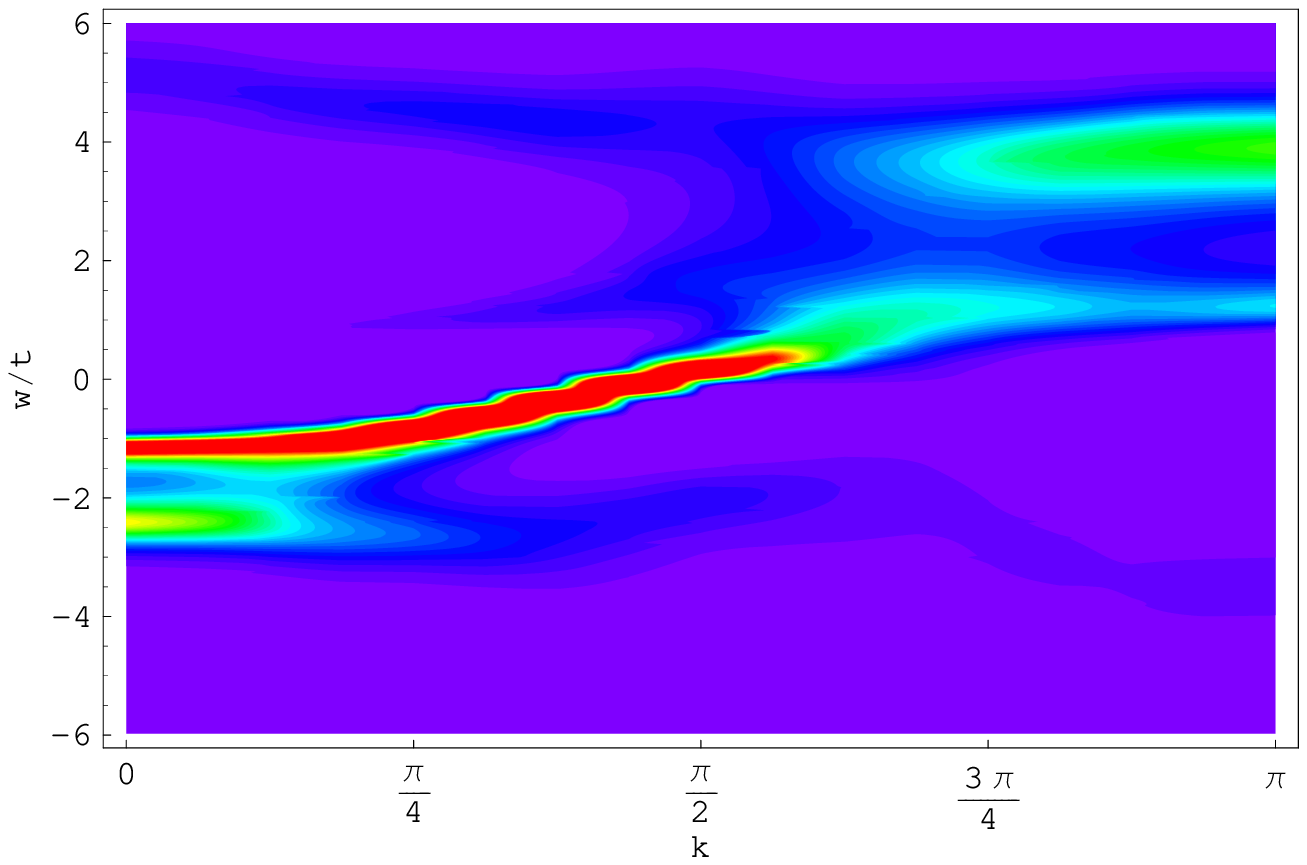}
\caption{Spectral function $A(\vec{k},\protect\omega )$ for (a) $N_{c}=2$,
(b) $N_{c}=12$, (c) $N_{c}=24$, and dispersion curve (bottom) for $N_{c}=24$
in the one-dimensional Hubbard model for $U/t=4$, $\protect\beta =5$, $n=0.89
$. }
\label{A_k_w_4.0_1.00_.00_nc=24.fig}
\end{figure}

As an application of CDMFT+QMC, we present in this section a study of the
appearance of spinon and holons as a function of system size at finite
temperature. The spectral function $A(\vec{k},\omega )$ and its dispersion
curve are calculated in the one-dimensional Hubbard model for $U/t=4$, $%
\beta =5$, $n=0.89$ with several sizes of cluster (obtained without
weighting) shown in Fig.~\ref{A_k_w_4.0_1.00_.00_nc=24.fig} to demonstrate the ability of CDMFT to reproduce highly
nontrivial physics in one-dimensional systems. For $N_{c}=2,$ $A(\vec{k}%
,\omega )$ has only one broad feature near $\vec{k}=0$ and $\pi $, while for $%
N_{c}=12$ it starts showing, near $\vec{k}=\pi $, continuous spectra that
are bounded by two sharp features. Near $\vec{k}=0$ the two features do not
show up clearly. For $N_{c}=24$ however, the spectral function shows the
separation of spinon and holon dispersions near both $\vec{k}=0$ and $\pi $, even if
the temperature $\beta =5$ is relatively large. 
These features are in agreement with recent QMC calculations
for the one-dimensional Hubbard
model~\cite{Matsueda:2005,Abendschein:2006}.
As $\vec{k}$ approaches $%
\vec{k}_{F}$, we loose the resolution necessary to separate the two spectra.
For $U/t=6$ we obtain a similar result, while at weak coupling ($U/t=2$) we
do not resolve the separation up to $L=24$. 


\section{Summary, conclusions and outlook}

\label{section6}

To summarize, we have studied the Hubbard model as an example of strongly
correlated electron systems using the Cellular Dynamical Mean-Field Theory
(CDMFT) with quantum Monte Carlo (QMC) simulations. The cluster problem may
be solved by a variety of techniques such as exact diagonalization (ED) and
QMC simulations. We have presented the algorithmic details of CDMFT with the
Hirsch-Fye QMC method for the solution of the self-consistently embedded
quantum cluster problem. We have used the one-dimensional half-filled
Hubbard model to benchmark the performance of CDMFT+QMC particularly for
small clusters by comparing with the exact results.
We have also calculated the single-particle Green's
functions and self-energies on small clusters to study the size dependence
of the results in one- and two-dimensions, and finally, we have shown that
spin-charge separation in one dimension can be studied with this approach
using reasonable cluster sizes. 

To be more specific, it has been shown that in one dimension, CDMFT+QMC with
two sites in the cluster is already able to describe with high accuracy the
evolution of the density as a function of chemical potential and the
compressibility divergence at the Mott transition, in good agreement
with the exact Bethe ansatz result. This presents an independent
confirmation of the ability of CDMFT to reproduce the compressibility
divergence with small clusters. In the previous tests with CDMFT+ED, some
sensitivity to the so-called distance function, had been noticed~\cite%
{CCKCK:2004}. This question does not arise with QMC. This is very
encouraging, since mean field methods would be expected to perform even
better as the dimensionality increases. We also looked at the cluster size
dependence of the Green's function $G(\vec{k},\tau ).$ It becomes smaller in
magnitude with increasing system size and the infinite cluster limit is
approached in the opposite way to that in finite size simulations, as was
observed before in another quantum cluster scheme (DCA)~\cite{MHJ:2000}.
With increasing $U$ the result on the smallest cluster rapidly approaches
that of the infinite size cluster. Large scattering rate and a positive
slope in the real part of the self-energy in one-dimension suggest that the
system is a non-Fermi liquid for all the parameters studied here.

In two-dimensions, a similar size dependence to the one-dimensional case is
found. At weak coupling a pseudogap appears only for large clusters in
agreement with the expectation that at weak coupling a large correlation
length (on a large cluster) is required to create a gap. At intermediate to
strong coupling, even the smallest cluster ($N_{c}=2\times 2$) accounts for
more than $95\%$ of the correlation effect in the single particle spectrum
of the infinite size cluster, (as measured by Eq.(\ref{eq170})). This is
consistent with our earlier study that showed indirectly that for $U$ equal
to the bandwidth or larger, short-range correlation effect (available in a
small cluster) starts to dominate the physics~\cite{KKSTCK:2005}. This
presents great promise that some of the important problems in strongly
correlated electron systems may be studied highly accurately with a
reasonable computational effort.

Finally, we have shown that CDMFT+QMC can describe highly nontrivial long
wavelength Luttinger liquid physics in one-dimension. More specifically, for 
$U=4$ and $\beta =5$ the separation of spinon and holon dispersions is clear
even for $N_{c}=24.$

Issues that can now be addressed in future work include that of the origin
of the pseudogap observed in hole-underdoped cuprates. Since the parent
compounds of the cuprates are Mott-Hubbard insulators, an understanding of
such an insulator and its evolution into a correlated metal upon doping is
crucial. In particular, CDMFT+QMC offers the possibility of calculating the
pseudogap temperature to compare with experiment. This has been successfully
done at intermediate coupling with TPSC~\cite{TKS:2005}, but at strong
coupling, quantum cluster approaches are needed. Single-site DMFT is not
enough since, for example, high resolution QMC study for the half-filled 2D
Hubbard model~\cite{MHSD:1995, PLH:1995} found two additional bands besides
the familiar Hubbard bands in the spectral function. These are apparently
caused by short-range spatial correlations that are missed in the
single-site DMFT. The search for a coherent understanding of the evolution
of a Mott insulator into a correlated metal by doping at finite temperature
has been hampered by the severe minus problem in QMC away from half-filling
and at low temperature. An accurate description of the physics at
intermediate to strong coupling with CDMFT+QMC with small clusters (as shown
in this paper) and modest sign problems in quantum cluster methods (as shown
in DCA~\cite{JMHM:2001}) give us the tools to look for a systematic physical
picture of the finite temperature pseudogap phenomenon at strong coupling. 

Another issue that can be addressed with CDMFT+QMC is that of the
temperature range over which spin-charge separation occurs in the
one-dimensional Hubbard model. In other words, at what temperature does
Luttinger Physics breaks down as a function of $U$, and when it breaks down
what is the resulting state? We saw in this paper that even a single peak in
the single-particle spectral weight does not immediately indicate a Fermi
liquid. 

Finally, one methodological issue. Two-particle correlation functions are
necessary to identify second order phase transitions by studying the
divergence of the corresponding susceptibilities. This can be done with 
DCA~\cite{MJPH:2004}. In the present paper, instead, we focused on one-particle
quantities such as the Green's function and related quantities.  In some
sense, quantum cluster methods such as CDMFT use irreducible quantities
(self-energy for one-particle functions and irreducible vertices for
two-particle functions) of the cluster to compute the corresponding lattice
quantities. Since the CDMFT is formulated entirely in real space and the
translational symmetry is broken at the cluster level, it appears extremely
difficult, in practice, to obtain two-particle correlation functions and
their corresponding irreducible vertex functions in a closed form like
matrix equations to look for instabilities. One way to get around this
problem is, as in DMFT, to introduce mean-field order parameters such as
antiferromagnetic and $d-wave$ superconducting orders, and to study if they
are stabilized or not for given parameters such as temperature and doping
level. In this way one can, for example, construct a complete phase diagram
of the Hubbard model, including a possible regime in which several phases
coexist. Zero-temperature studies with CDMFT+ED~\cite{KCCKSKT:2005} and with
the Variational Cluster Approximation~\cite{SLMT:2005,HAAP:2005} have
already been performed along these lines. 

\acknowledgements We thank S. Allen, C. Brillon, M. Civelli, A. Georges, S.
S. Kancharla, V. S. Oudovenko, O. Parcollet, and D. S\'{e}n\'{e}chal for
useful discussions, and especially S. Allen for sharing his maximum entropy
code. Computations were performed on the Elix2 Beowulf cluster and on the
Dell cluster of the RQCHP. The present work was supported by NSERC (Canada),
FQRNT (Qu\'{e}bec), CFI (Canada), CIAR, the Tier I Canada Research Chair
Program (A.-M.S.T.) and the NSF under grant DMR-0096462 (G.K.).



\begin{thebibliography}{99}
\bibitem{Dagotto:1994} E. Dagotto, Rev. Mod. Phys. \textbf{66}, 763 (1994).

\bibitem{HTJPK:1998} M. H. Hettler, A. N. Tahvildar-Zadeh, M. Jarrell, T.
Pruschke, and H. R. Krishnamurthy, Phys. Rev. B \textbf{58}, R7475 (1998).

\bibitem{SPP:2000} D. S\'en\'echal, D. Perez, and M. Pioro-Ladri\`ere, Phys.
Rev. Lett. \textbf{84}, 522 (2000).

\bibitem{LK:2000} A. I. Lichtenstein and M. I. Katsnelson, Phys. Rev. B 
\textbf{62}, R9283 (2000).

\bibitem{KSPB:2001} G. Kotliar, S. Savrasov, G. Pallson, and G. Biroli,
Phys. Rev. Lett. \textbf{87}, 186401 (2001).

\bibitem{P:2003} M. Potthoff, Eur. Phys. J. B \textbf{32}, 429 (2003); Phys.
Rev. Lett. \textbf{91}, 206402 (2003).

\bibitem{MJPH:2004} T. Maier, M. Jarrell, T. Pruschke, and M. H. Hettler,
Rev. Mod. Phys. \textbf{77}, 1027 (2005) .

\bibitem{GK:1992} A. Georges and G. Kotliar, Phys. Rev. B \textbf{45}, 6479
(1992).

\bibitem{Jarrell:1992} M. Jarrell, Phys. Rev. Lett. \textbf{69}, 168 (1992).

\bibitem{GKKR:1996} A. Georges, G. Kotliar, W. Krauth, and M. J. Rozenberg,
Rev. Mod. Phys. \textbf{68}, 13 (1996).

\bibitem{TS:1999} T. Timusk and B. Statt, Rep. Prog. Phys. {\bf 62}, 61 (1999);
M. R. Norman, D. Pines and C. Kallin, Adv. Phys. \textbf{54}, 715 (2005).

\bibitem{PBK:2004} O. Parcollet, G. Biroli, and G. Kotliar, Phys. Rev. Lett. 
\textbf{92}, 226402 (2004).

\bibitem{HF:1986} J. Hirsch and R. Fye, Phys. Rev. Lett. \textbf{56}, 2521
(1986).

\bibitem{KCCKSKT:2005} S. S. Kancharla, M. Civelli, M. Capone, B. Kyung, D. S%
\'{e}n\'{e}chal, G. Kotliar, A.-M.S. Tremblay, cond-mat/0508205.

\bibitem{BKK:2003} C. Bolech, S. S. Kancharla, and G. Kotliar, Phys. Rev. B 
\textbf{67}, 075110 (2003).

\bibitem{CCKCK:2004} M. Capone, M. Civelli, S. S. Kancharla, C. Castellani,
and G. Kotliar, Phys. Rev. B \textbf{69}, 195105 (2004).

\bibitem{KKSTCK:2005} B. Kyung, S. S. Kancharla, D. S\'{e}n\'{e}chal, A. -M. S. Tremblay,
M. Civelli, and G. Kotliar, Phys. Rev. B \textbf{73}, 165114 (2006).

\bibitem{SK:2005} Tudor D. Stanescu and Gabriel Kotliar, cond-mat/0508302.

\bibitem{DS:2003} R.R. dos Santos, Brazilian Journal of Physics \textbf{33}
36 (2003).

\bibitem{Hirsch:1983} J. Hirsch, Phys. Rev. B \textbf{28}, 4059 (1983).

\bibitem{Comment:Deltafunction} $\delta _{l,l^{\prime }+1}$ (not $\delta
_{l,l^{\prime }}$) is required in order to reproduce the correct fermionic
Green's function from the determinant of $G_{\sigma ,\{s\}}^{-1}$. The
antiperiodicity comes from the antiperiodic boundary condition of the
fermionic Green's function under a shift of $\beta $. See R. Blankenbecler,
D. J. Scalapino, and R. L. Sugar, Phys. Rev. D \textbf{24}, 2278 (1981) and 
\cite{Hirsch:1983}.

\bibitem{JG:1996} M. Jarrell and J. Gubernatis, Physics Reports \textbf{269}%
, 133 (1996).

\bibitem{Allen:2005} S. Allen, private communications.

\bibitem{OK:2002} V. S. Oudovenko and G. Kotliar, Phys. Rev. B \textbf{65},
075102 (2002).

\bibitem{JMHM:2001} M. Jarrell, T. Maier, C. Huscroft, and S. Moukouri,
Phys. Rev. B \textbf{64}, 195130 (2001).

\bibitem{KSHOPM:2005} G. Kotliar, S. Y. Savrasov, K. Haule, V. S. Oudovenko,
O. Parcollet, and C.A. Marianetti, cond-mat/0511085.

\bibitem{MPI} W. Gropp, E. Lusk, and A. Skjellum, 1999, \textit{Using MPI:
Portable Parallel Programming with the Message Passing Interface}, 2nd ed.
(The MIT Press).

\bibitem{Comment:1D} For $\beta=20$ the kink near $\mu=1.45$ is rounded.

\bibitem{LW:1968} E. H. Lieb and F. Y. Wu, Phys. Rev. Lett. \textbf{20},
1445 (1968).

\bibitem{PK:2005} O. Parcollet and G. Kotliar, unpublished.

\bibitem{CCKPK:2005} M. Civelli, M. Capone, S. S. Kancharla, O. Parcollet,
and G. Kotliar, Phys. Rev. Lett. \textbf{95}, 106402 (2005).

\bibitem{MHJ:2000} S. Moukouri, C. Huscroft, and M. Jarrell,
cond-mat/0004279.

\bibitem{BK:2002} G. Biroli and G. Kotliar, Phys. Rev. B \textbf{65}, 155112
(2002).

\bibitem{MJ:2002} T. Maier and M. Jarrell, Phys. Rev. B \textbf{65},
041104(R) (2002).

\bibitem{AMJ:2005} K. Aryanpour, T. Maier, and M. Jarrell, Phys. Rev. B 
\textbf{71}, 037101 (2005).

\bibitem{Comment:Weighting} We believe that with appropriate weighting the
convergence can be much faster than linear in $1/L$ at large clusters.

\bibitem{Anderson:1987} P. W. Anderson, Science \textbf{235}, 1196 (1987).

\bibitem{TKS:2005} A. -M. S. Tremblay, B. Kyung, and D. S\'{e}n\'{e}chal,
Fizika Nizkikh Temperatur (Low Temperature Physics), \textbf{32}, 561 (2006).

\bibitem{Comment:2D} Because open boundary conditions are imposed in the
cluster, an odd number of linear size such as 3 and 5 can be used without
causing any irregularity. Although an open cluster breaks the full lattice
translational symmetry at the cluster level (which is eventually restored by
Eq.~\ref{eq40}), here it works to our advantage.

\bibitem{Vilk:1997} Y.~ Vilk and A.-M. Tremblay, J. Phys I (France) \textbf{7%
}, 1309 (1997).

\bibitem{Vilk:1996} Y.~ Vilk and A.-M. Tremblay, Europhys. Lett. \textbf{33}%
, 159 (1996).

\bibitem{KGT:2005} B. Kyung, A. Georges, and A. -M. S. Tremblay,
         cond-mat/0508645. 

\bibitem{CSSBDJ:2002} R. Claessen, M. Sing, U. Schwingenschloegl, P. Blaha,
M. Dressel, and C.S. Jacobsen, Phys. Rev. Lett. \textbf{88}, 096402 (2002).

\bibitem{Voit:1995} For a review, see J. Voit, Rep. Prog. Phys. \textbf{58},
977 (1995).

\bibitem{BGJ:2004} H. Benthien, F. Gebhard, and E. Jeckelmann, Phys. Rev.
Lett. \textbf{92}, 256401 (2004).

\bibitem{ZAHS:1998} M. G. Zacher, E. Arrigoni, W. Hanke, and J. R.
Schrieffer, Phys. Rev. B \textbf{57}, 6370 (1998).

\bibitem{Matsueda:2005} H. Matsueda, N. Bulut, T. Tohyama, and S. Maekawa,
         Phys. Rev. B \textbf{72}, 075136 (2005).

\bibitem{Abendschein:2006} A. Abendschein and F. F. Assaad,
         cond-mat/0601222.

\bibitem{MHSD:1995} A. Moreo, S. Hass, A. W. Sandvik, and E. Dagotto, Phys.
Rev. B \textbf{51}, 12 045 (1995).

\bibitem{PLH:1995} R. Preuss, W. von der Linden, and W. Hanke, Phys. Rev.
Lett. \textbf{75}, 1344 (1995).

\bibitem{SLMT:2005} D.~S\'{e}n\'{e}chal, P.-L. Lavertu, M.-A. Marois and
A.-M.~S. Tremblay, Phys. Rev. Lett. \textbf{94}, 156404 (2005).

\bibitem{HAAP:2005} W.~ Hanke, M.~ Aichhorn, E.~ Arrigoni and M.~ Potthoff, 
\emph{\bibinfo{title}{Correlated band structure and the ground-state phase
diagram in high-$T_c$ cuprates}} (2005), \eprint{cond-mat/0506364}.

\end{thebibliography}
\end{document}